\documentclass[prb,aps,showpacs,floatfix,floats,nobalancelastpage,twocolumn]{revtex4}
\usepackage{amsmath, eufrak}
\usepackage{graphicx}
\usepackage{amsmath}
\usepackage{color}
\usepackage{dcolumn}
\usepackage{bm}
\usepackage{subfigure}
\usepackage{ulem}

\def\etal{~\textit{et~al}}
\def\ra{\rangle}
\def\la{\langle}

\def\Hc{{\rm H.c.}}
\def\bk{{\bf k}}
\def\br{{\bf r}}
\def\bR{{\bf R}}
\def\bfm{{\bf m}}
\def\hbx{{\hat{\bf x}}}
\def\hby{{\hat{\bf y}}}
\def\cK{{\cal K}}
\def\cU{{\cal U}}

\begin{document}

\title{Possible realization of the Exciton Bose Liquid phase in a hard-core boson model with ring-only exchange interactions}

\author{Tiamhock Tay}
\author{Olexei I. Motrunich}

\affiliation{Department of Physics, California Institute of Technology, Pasadena, CA 91125}

\date{\today}


\begin{abstract}
We investigate a hard-core boson model with ring-only exchanges on a square lattice, where a $K_1$ term acts on 1$\times$1 plaquettes and a $K_2$ term acts on 1$\times$2 and 2$\times$1 plaquettes, with a goal of realizing a novel Exciton Bose Liquid (EBL) phase first proposed by Paramekanti\etal~[Phys.~Rev.~B {\bf 66}, 054526 (2002)].    We construct Jastrow-type variational wave functions for the EBL, study their formal properties, and then use them as seeds for a projective Quantum Monte Carlo study.   Using Green's Function Monte Carlo, we obtain an unbiased phase diagram which at half-filling reveals CDW for small $K_2$, valence bond solid for intermediate $K_2$, and possibly for large $K_2$ the EBL phase.  Away from half-filling, the EBL phase is present for intermediate $K_2$ and remains stable for a range of densities below 1/2 before phase separation occurs at lower densities.
\end{abstract}
\maketitle

\section{Introduction}

The Exciton Bose Liquid (EBL) theory by Paramekanti\etal~ proposed a critical bosonic phase which shows remarkable resemblance to electrons in a metal.\cite{Paramekanti2002}  For this novel quantum phase, the presence of loci of ``Bose surface'' in the Brillouin zone closely parallels the Fermi surface for fermions, and allows the Bose system to share many characteristics normally associated with fermions.\cite{Sachdev2002}  Paramekanti\etal\ showed that due to the gapless lines of excitations, EBL is a critical (power law) compressible 2D quantum phase with uncondensed bosons and contains continuously varying exponents.  Their striking proposal stimulated a number of works seeking to establish the stability of the EBL phase in bosonic models with ring exchange interactions.\cite{Sandvik2002,Melko2004,Rousseau2004,Rousseau2005}  However, these studies found that the EBL is not realized in the hard-core boson model on the square lattice with ring exchanges on elementary plaquettes.  Instead, such ring interactions favor a $(\pi, \pi)$ charge density wave (CDW) in the half-filled case, while away from half-filling they induce strong tendencies to phase separation.

\begin{figure}
  \centering
  \includegraphics[width=\columnwidth]{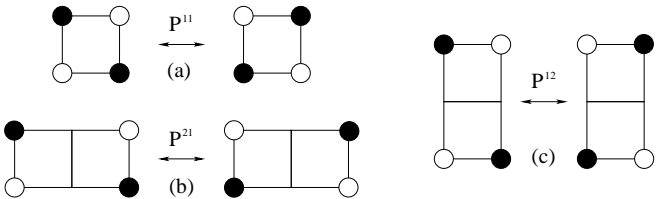}
  \caption{The ring exchange interactions for (a) 1$\times$1, (b) 2$\times$1 and (c) 1$\times$2 plaquettes, which hop two bosons on opposite corners of a plaquette onto the two remaining vacant corners.}
  \label{fig:ring}
\end{figure}

Motivated by the unusual EBL properties, our study focuses on a search for this unconventional quantum phase in simple bosonic models.   A candidate model requires additional interactions for suppressing the charge order.  One choice that comes to mind might be to introduce second nearest neighbor repulsion.   In this paper, we instead adopt a different route where we consider additional ring exchanges that frustrate the CDW tendencies of the elementary ring exchanges.   We define plaquette exchange operators
\begin{equation}
  P^{mn}_{\bf r} = b_{\bf r}^\dagger~ b_{{\bf r}+m{\bf \hat{x}}}~ b_{{\bf r}+m{\bf \hat{x}}+n{\bf \hat{y}}}^\dagger~ b_{{\bf r}+n{\bf \hat{y}}} + \Hc ~,
\end{equation}
where $b_{\bf r}$ annihilates a boson on a site ${\bf r}$, and ${\bf \hat{x}}, {\bf \hat{y}}$ are the unit vectors on the square lattice.  The Hamiltonian is
\begin{equation}
  \hat{H} = -K_1\sum_{\bf r} P^{11}_{\bf r} - K_2 \sum_{\bf r}\left(P^{12}_{\bf r} + P^{21}_{\bf r}\right) ~.
\end{equation}
Figure~\ref{fig:ring} illustrates the action of these ring exchanges on hoppable plaquettes.    The original ring model proposed in Ref.~\onlinecite{Paramekanti2002} and studied numerically in Refs.~\onlinecite{Sandvik2002,Melko2004,Rousseau2004,Rousseau2005} corresponds to $K_2=0$.    To see how the present $K_1$-$K_2$ model may stabilize the EBL phase, we first note that the ($\pi$,$\pi$) CDW in the $K_1$-only model results from having a large number of basis states connected to the perfect ($\pi$,$\pi$) CDW configuration.   However, the $K_2$ terms would be completely inoperative in such a CDW.  Furthermore, the $P^{12}$ and $P^{21}$ ring exchanges by themselves would favor different charge orderings and not compatible with each other.  Thus, the $K_2$ ring terms compete with the $K_1$ terms and with each other, making the liquid phase with no charge order more likely.

The present $K_1$-$K_2$ model has the same lattice symmetries and boson number conservation on each row and column as the original ring model of Ref.~\onlinecite{Paramekanti2002}.    From the outset, we define our Hilbert space as the sector with equal number of bosons on each row and column.   Note that this restriction does not preclude phase separation (PS); in fact, we shall see that PS does occur at low densities within our restricted Hilbert space.    For non-negative $K_1$ and $K_2$ values, the Hamiltonian does not have a sign problem and allows an unbiased study of the system using Quantum Monte Carlo methods.  Although the Stochastic Series Expansion is the method of choice for simulating large lattices, it has not been applied to ring-only hard-core boson models due to implementation issues.\cite{Sandvik2002}    We instead use the Green's Function Monte Carlo (GFMC) approach with full bias control as described in Ref.~\onlinecite{Buonaura1998}, which allows us to obtain exact ground state properties for moderately sized systems up to $12 \times 12$ in this work.    Without loss of generality, we set $K_1 = 1$ and vary $K_2\geq 0$ in the study.

Our main results for the phases of the model are summarized in Figs.~\ref{fig:GFMC_phase_diagram} and \ref{fig:completePD}.  First, at half-filling, our intuition that the $K_2$ should suppress the charge order is indeed borne out, and the CDW disappears already for moderate $K_2$.  Somewhat surprisingly, this does not stabilize the EBL right away but instead drives the system into a columnar Valence Bond Solid (VBS), while the EBL is tentatively stabilized only for quite large $K_2$ terms.  On the other hand, away from half-filling, moderate $K_2$ already produce stable EBL phase.

The paper is organized as follows.  In Sec.~\ref{sec:VMC}, we first construct good trial wave functions for the EBL phase and study their formal properties using Variational Monte Carlo (VMC), followed by an energetics study to determine the variational phase diagram.  Using the optimal trial states as starting point for the GFMC projection in Sec.~\ref{sec:GFMC}, we compute density, plaquette, and bond structure factors in the ground states, followed by finite size scaling to determine the phases.  Our phase diagram reveals the ($\pi$,$\pi$) CDW order at small $K_2$, VBS order for intermediate $K_2$, and possibly for large $K_2$ the novel EBL phase. We perform detailed comparison of the numerical results with the EBL theory.   In Sec.~\ref{sec:general_densities}, we extend the search for the EBL to densities less than half at intermediate $K_2$, and find that the EBL liquid is stable for $1/3 \lesssim \rho < 1/2$ while phase separation occurs at lower densities.  In Sec.~\ref{sec:conclusion}, we conclude with a discussion of possible further studies.
For the benefit of readers,  Appendix~\ref{app:EBLprecis} summarizes the results of the EBL theory from Refs.~\onlinecite{Paramekanti2002, Xu2005, Xu2007, Balents2005} relevant for our numerical work.  In Appendix~\ref{app:parton2EBL}, we offer a parton-gauge perspective of the EBL theory.

\section{Variational Study at $\rho=1/2$}\label{sec:VMC}

\subsection{Formal properties of the EBL wave function}\label{sec:VMC_formal}

In this section, we study the formal properties of the EBL wave function. To provide motivation for the wave function, we first consider a quantum rotor version of the model with elementary 1$\times$1 ring exchanges,
\begin{eqnarray}
  \hat{H}_{\rm rotor} &=& - K \sum_{\bf r} \cos\left( \phi_{\bf r} - \phi_{\bf r+\hat{x}} + \phi_{\bf r+\hat{x}+\hat{y}} - \phi_{\bf r+\hat{y}} \right) \nonumber\\
  &+& \frac{U}{2} \sum_{\bf r} \left( n_{\bf r} - \bar{n} \right)^2,
\end{eqnarray}
where the phase $\phi_{\bf r}$ and the boson number $n_{\bf r}$ are canonically conjugate.  In the EBL theory, the cosine in the ring term is expanded to quadratic order (this approximation is valid in the stable ``spin-wave phase'' with no topological defects).  The resulting set of coupled harmonic oscillators can be diagonalized in momentum space, which leads to the following,
\begin{eqnarray}
  \hat{H}_{\rm SW} &=& \sum_{\bf q} \left( \frac{U}{2}~ n_{\bf q}~n_{-\bf q} + \frac{\omega_{\bf q}^2}{2U}~ \phi_{\bf q}~\phi_{-\bf q} \right), \label{eq:hamiltonian_SW}\\
  \omega_{\bf q} &=& 4 \sqrt{U K} \left\vert \sin\left(\frac{q_x}{2}\right) \sin\left(\frac{q_y}{2}\right) \right\vert ~.
\end{eqnarray}
We will loosely refer to $\hat{H}_{\rm SW}$ as the ``spin-wave'' Hamiltonian.  Writing its ground state in the $n$ variables and then restricting to $n_{\bf r} \in \{0,1\}$, we obtain a valid hard-core boson wave function in the convenient Jastrow-type form that can be implemented easily in VMC,
\begin{eqnarray}
  \Psi_{\rm EBL} &\propto& \exp\left[ -\frac{1}{2}\sum_{{\bf r},{\bf r}'} u({\bf r} - {\bf r}')~n_{\bf r} n_{{\bf r}'} \right],\label{eq:PsiEBL}\\
  u({\bf r}) &=& \frac{1}{L^2} \sum_{\bf q} \frac{W~ e^{i {\bf q} \cdot {\bf r}}}{4\left\vert \sin(q_x/2) \sin(q_y/2) \right\vert}.\label{eq:EBL_pseudo_potential}
\end{eqnarray}  
In the spin wave theory, $W = \sqrt{U/K}$, while here it serves as a variational parameter.  For the $K_1$-$K_2$ model, $W$ becomes a ${\bf q}$-dependent function with two parameters.  In this section, we will focus on the single-parameter EBL wave function to illustrate properties of such variational states and what can happen with them.

\begin{figure}
  \centering
  \includegraphics[scale=0.8]{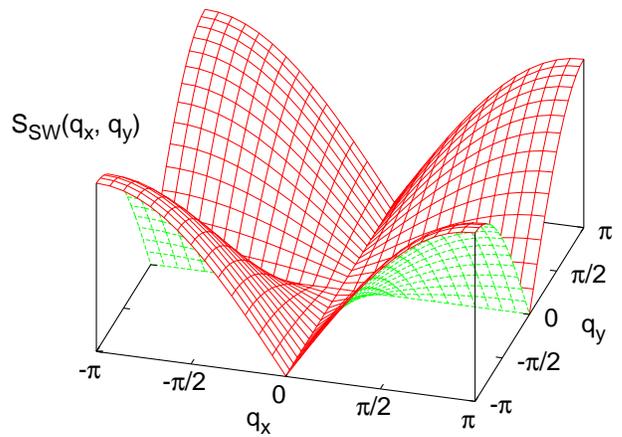}
  \caption{Density structure factor $S_{\rm SW}(q_x,q_y)$, Eq.~(\ref{eq:SqxqySW}), for the ground state of the spin-wave Hamiltonian, Eq.~(\ref{eq:hamiltonian_SW}).  The characteristic ``cross'' formed by the singular lines $q_x = 0$ or $q_y = 0$ is a distinctive feature of the EBL phase.
}
  \label{fig:SpinWave_SF_k05}
\end{figure}

To characterize the phases realized in the EBL wave function, we measure the density structure factor
\begin{equation}
  S(q_x,q_y) = \frac{1}{L^2} \sum_{{\bf r}, {\bf r}'} e^{i {\bf q} \cdot ({\bf r} - {\bf r}')} \la n_{\bf r}n_{{\bf r}'} - \bar{n}^2 \ra  ~.\label{eq:Sqxqy}
\end{equation}
The density structure factor of the ground state for the spin-wave Hamiltonian in Eq.~(\ref{eq:hamiltonian_SW}) is given by
\begin{equation}
  S_{\rm SW}(q_x,q_y) = \frac{2}{W}\left\vert \sin\left(\frac{q_x}{2}\right)\sin\left(\frac{q_y}{2}\right)\right\vert.\label{eq:SqxqySW}
\end{equation}
At any fixed $q_y$, $S_{\rm SW}(q_x,q_y)$ vanishes for small $q_x$ as $C(q_y)\vert q_x \vert$, with further $C(q_y) \sim |q_y|$ as $q_y \rightarrow 0$.   This gives the characteristic ``cross'' shown in Fig.~\ref{fig:SpinWave_SF_k05} which is a signature of the compressibility of the EBL, and is related to the gaplessness of $\omega_{\bf q}$ along the loci $q_x=0$ or $q_y=0$.  The latter is a consequence of the conservation of boson number along each row and each column of the lattice.   To identify possible realization of the EBL phase, we monitor the long wavelength behavior of the density structure factor in addition to the absence of Bragg peaks in all structure factor measurements made in this paper.

\begin{figure}
  \centering
  \includegraphics[width=\columnwidth]{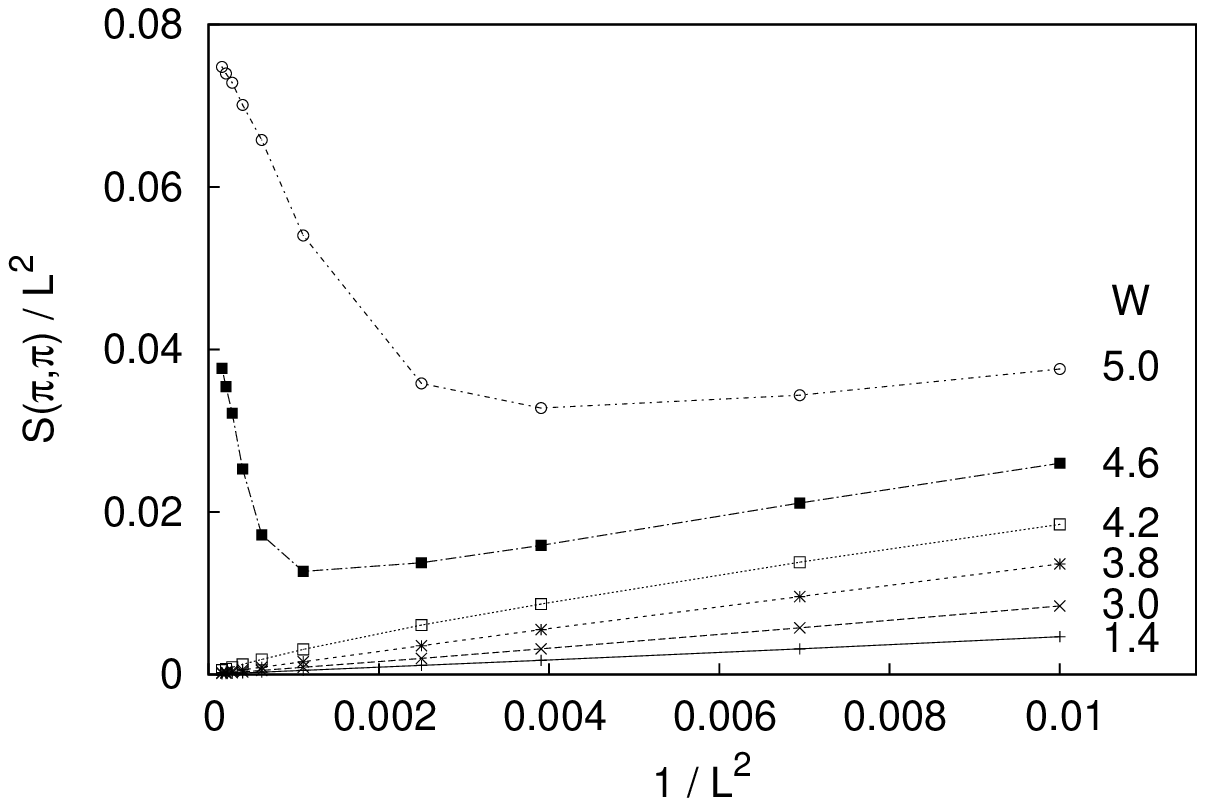}\\
  \includegraphics[width=\columnwidth]{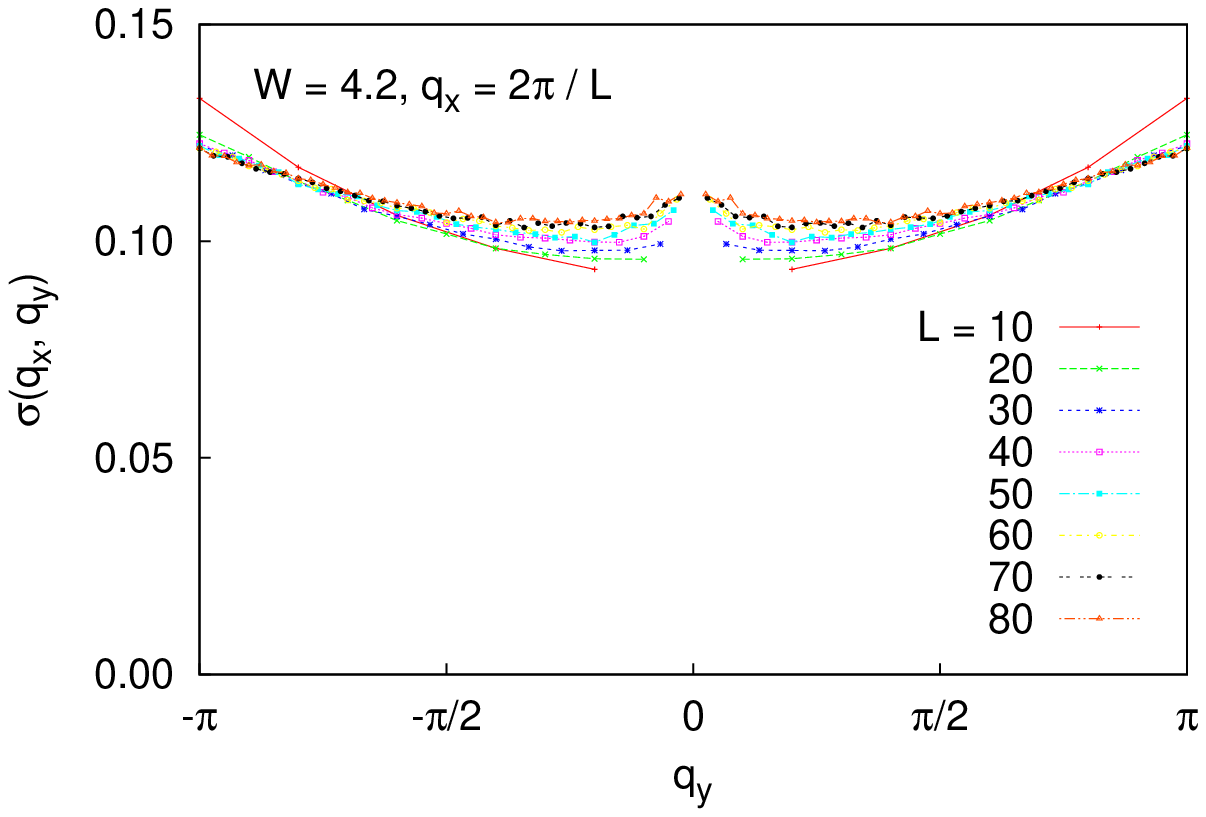}\\
  \includegraphics[width=\columnwidth]{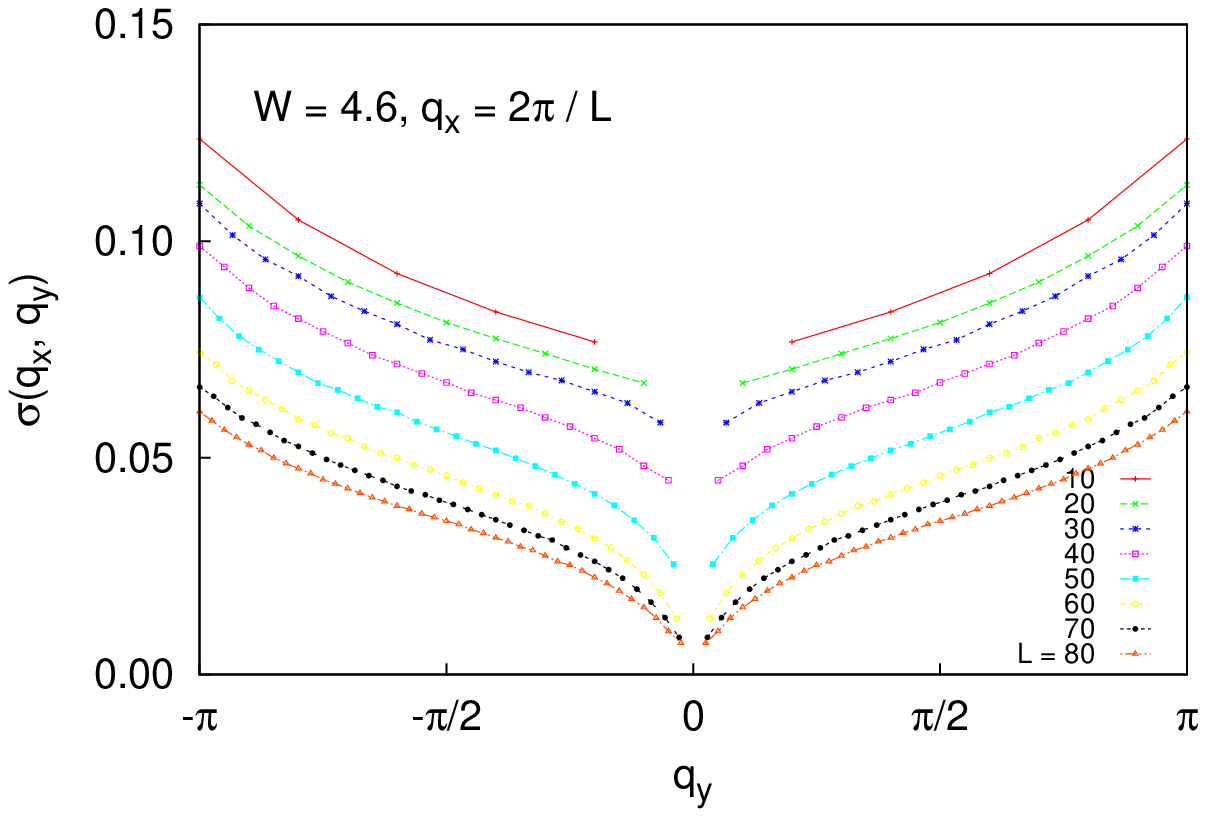}
  \caption{(color online).  VMC study of the EBL wave function in Eq.~(\ref{eq:PsiEBL}) with one parameter $W$ defined by Eq.~(\ref{eq:EBL_pseudo_potential}).  Top: Finite size scaling of the density structure factor $S(\pi,\pi)$.  The wave function undergoes a transition at a critical value $W_c \approx 4.4$, which separates the EBL phase at low $W$ from the ($\pi$,$\pi$) CDW at larger $W$.  Middle and bottom: $\sigma(q_x \!=\! 2\pi/L, q_y)$, Eq.~(\ref{sigma}), which gives normalized slopes of the density structure factor near the cross, plotted against $q_y$ for lattice length $L$ = 10 to 80.  Middle: Results for $W=4.2$ show long wavelength EBL characteristics where these normalized slopes approach finite values [apparently close to $\sigma_{\rm SW} = 1/(2W)$ indicated with a dotted line].  Bottom: Results for $W=4.6$ show a downward renormalization of the slopes, which is similar to a gapped state.  The critical values estimated from both the order parameter study and the detailed study of the cross agree, i.e., $W_c\approx 4.4$.}
  \label{fig:VMC_wavefunction_transition}
\end{figure}

For sufficiently strong interaction $U$ (in particular, for hard-core model) at half-filling, the spin-wave approximation no longer holds.     Proliferation of topological defects results in the ($\pi$,$\pi$) CDW instability found in earlier Quantum Monte Carlo studies.\cite{Paramekanti2002, Sandvik2002, Melko2004, Rousseau2004, Rousseau2005}  Remarkably, we find that the single-parameter EBL wave function Eq.~(\ref{eq:PsiEBL}) is able to realize both the EBL and the CDW phase.     As we increase $W$ in the half-filled system, the wave function undergoes a phase transition at a critical value $W_c \approx 4.4$ where the $(\pi,\pi)$ charge order develops.  This is analyzed in the top panel of Fig.~\ref{fig:VMC_wavefunction_transition} using finite size scaling of $S(\pi,\pi)/L^2$, which vanishes as $1/L^2$ in the absence of the order for $W < W_c$ and approaches a finite value in the presence of the order for $W > W_c$.  The non-monotonic $L$ dependence of this CDW order parameter for fixed $W > W_c$ is somewhat unusual but appears to be a property of such wave functions, perhaps indicative of some long crossovers in the system.

Next we examine the long wavelength behavior of the density structure factor near the characteristic ``cross''.  Figure~\ref{fig:VMC_wavefunction_transition} also shows the ratio 
\begin{eqnarray}
  \sigma(q_x, q_y) \equiv \frac{S(q_x, q_y)}{4 |\sin(q_x/2) \sin(q_y/2)|}
  \label{sigma}
\end{eqnarray}
evaluated at the smallest $q_x = q_{\rm min} = 2\pi/L$ as a function of $q_y$ on lattices with length $L$ between 10 and 80.  For $W=4.2$ in the middle panel, the ratio shows some deviation from $\sigma_{\rm SW} = 1/(2W)$ but it clearly renormalizes towards finite values as expected in the EBL theory.  When plotted in the full Brillouin zone, the VMC density structure factor looks essentially like Fig.~\ref{fig:SpinWave_SF_k05}.

On the other hand, for $W=4.6$ in the bottom panel, a strong downward renormalization of the ratio is observed for all $q_y$, in particular near $q_y=0$.  Such behavior is similar to a Mott insulator, where the density structure factor is nonsingular and hence has cuts ($q_x\rightarrow 0, q_y$) with vanishing slopes.  Thus, the contrasting long-wavelength behaviors of $S(q_x,q_y)$ independently confirm a phase transition in the wave function near $W_c\approx 4.4$. 

To summarize, the above wave function with one variational parameter can realize either the EBL liquid phase or the $(\pi,\pi)$ charge order and thus can alert us about CDW tendencies in the system.   A note of caution is appropriate here.  Our GFMC study in later Sec.~\ref{sec:GFMC} shows that the formal wave function study of the present section does not always capture a full physics of the problem.  Namely, as we will discuss in Sec.~\ref{subsec:VMCfailure}, the wave function itself may be in the liquid phase, while the full EBL theory with the same effective $K/U$ is already unstable.  Nevertheless, our formal wave function study clearly has its own merits.  For instance, it alerts us to the possibility of complex crossovers with the system size and that the order may be weak and not apparent on short scales, but may still appear on longer scales.  It also teaches us to look at the long-wavelength behavior for signs of instabilities.

\subsection{$K_1$-$K_2$ energetics study with one-parameter EBL wave function}

\begin{figure}
  \includegraphics[width=\columnwidth]{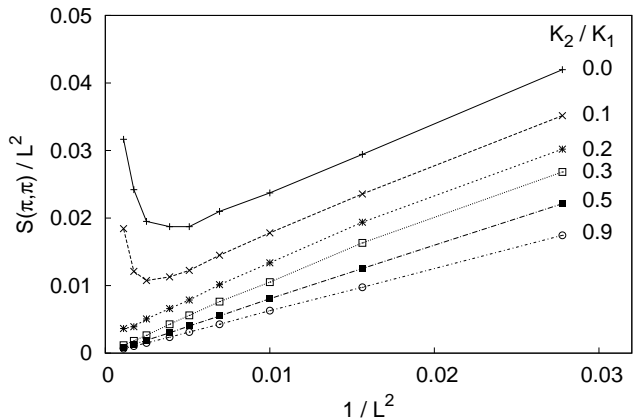}
  \caption{Finite size scaling of the VMC density structure factor $S(\pi,\pi)$ obtained in the energetics study using optimized $W$ for each $K_2$ in the Hamiltonian and each system size $L$.}
  \label{fig:VMC_SF_scaling}
\end{figure}

\begin{figure}
  \centering
  \includegraphics[width=\columnwidth]{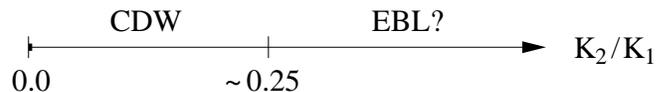}
  \caption{Variational phase diagram for the $K_1$-$K_2$ model on the half-filled square lattice, based on the single-parameter trial wave function.  Refer to Fig.~\ref{fig:GFMC_phase_diagram} for a revised phase diagram based on unbiased GFMC results.}
  \label{fig:vmc_phase_diagram}
\end{figure}

Let us turn to the energetics study using the above one-parameter wave function.  For each value of $K_2$ and system size $L$ we find the optimal variational parameter $W$ (later Fig.~\ref{fig:comparison_of_energies_SF_for_6x6_latt} compares the trial energies with the exact diagonalization values on the 6$\times$6 lattice).  In Fig.~\ref{fig:VMC_SF_scaling}, finite size scaling of the corresponding structure factor $S(\pi,\pi)$ shows the presence of the CDW order for small $K_2$ and the absence for $K_2 \geq 0.3$.  Thus, the VMC study suggests that the EBL phase could be stabilized even with quite weak $K_2$ ring interactions.   Allowing an additional variational parameter corresponding to $K_2$ in the spin-wave Hamiltonian does not modify this conclusion.   Figure~\ref{fig:vmc_phase_diagram} summarizes the VMC results obtained for the $K_1$-$K_2$ ring model at half-filling.   In the following section, our GFMC simulation reveals another phase in the intermediate $K_2$ region.   We will present a revised phase diagram based on unbiased GFMC results in Sec.~\ref{sec:GFMC_half_filling}, and will discuss the failure of the VMC later in Sec.~\ref{subsec:VMCfailure}.

\section{Unbiased GFMC Study at $\rho=1/2$}\label{sec:GFMC}

In this section, we perform a Green's function Monte Carlo study which, being unbiased, provides an important check on the VMC results.  In the GFMC approach, an initial trial wave function is projected onto the ground state via repeated application of a projector which eventually filters out the excited states:
\begin{equation}
  \vert \psi_{n+1}\ra = [1 - (\hat{H}-E_0)\delta\tau] \vert \psi_n \ra.
\end{equation}
Here $E_0$ is a parameter chosen close to the ground state energy, and $\delta\tau$ is a ``time step'' chosen to ensure dominance of the ground state and positiveness of the projector, which then allows Monte Carlo calculations without a sign problem.     Operator expectation values are evaluated using stochastically sampled ground states which generally requires the so-called GFMC ``forward walking'' technique, and we implement this using the bias-controlled scheme described in Ref.~\onlinecite{Buonaura1998}.  We refer the reader to the literature for more details on the GFMC.\cite{Hetherington1984, Trivedi1990, Runge1992, Buonaura1998}

To identify the nature of the true ground states, we measure the density structure factor $S(q_x,q_y)$ defined in Eq.~(\ref{eq:Sqxqy}) as well as the following plaquette structure factor
\begin{equation}
  P(q_x,q_y) = \frac{1}{L^2}\sum_{\bf r,r'} e^{i{\bf q}\cdot({\bf r}-{\bf r'})}\la (P^{11}_{\bf r})^2 ~(P^{11}_{\bf r'})^2 \ra ~,\label{Eq:Pqxqy}
\end{equation}
where $(P^{11}_{\bf r})^2$ equals $1$ if the 1$\times$1 plaquette is ``hoppable'' and $0$ otherwise.  While quantitatively different from the off-diagonal $P^{11}_{\bf r}$ plaquette structure factor used in Ref.~\onlinecite{Sandvik2002}, the operator $(P^{11}_{\bf r})^2$ defined here is easier to implement in the GFMC and it gives qualitatively the same access to bond-solid--type phases.  To better discriminate between plaquette and bond orders, we also measure the following bond structure factor
\begin{eqnarray}
  B_\alpha(q_x,q_y) &=& \frac{1}{L^2}\sum_{\bf r,r'} e^{i{\bf q}\cdot({\bf r}-{\bf r'})}\la (B_{\bf r}^\alpha)^2 ~(B_{\bf r'}^\alpha)^2 \ra ~,\label{Eq:Bqxqy}
\end{eqnarray}
where $B_{\bf r}^\alpha = b_{\bf r}^\dagger b_{\bf r+\hat{\alpha}} + b_{\bf r+\hat{\alpha}}^\dagger b_{\bf r}$ and $\alpha \in \{\hat{x},\hat{y}\}$; thus, $(B_{\bf r}^\alpha)^2$ is 1 if the bond is ``hoppable'' and 0 otherwise, and is again easy to implement in the GFMC.

\subsection{Test of our GFMC setup}

\begin{figure}
  \includegraphics[width=\columnwidth]{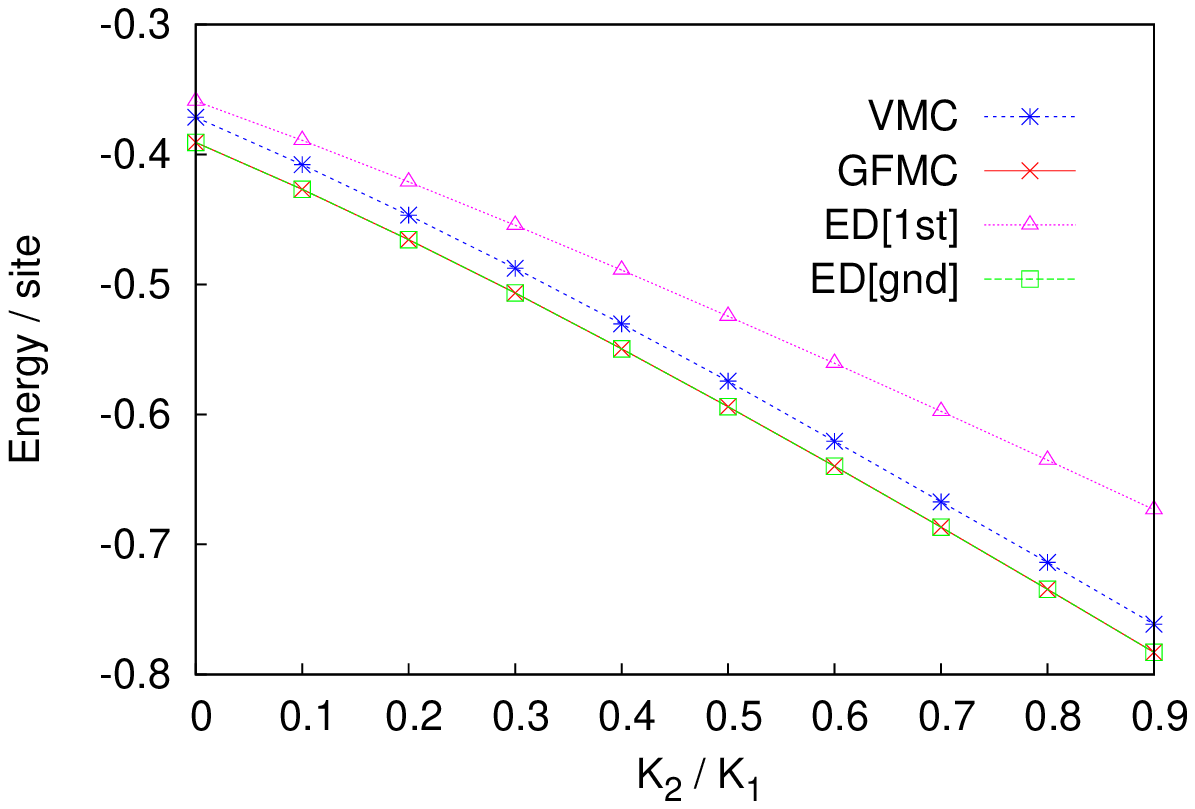}\\
  \includegraphics[width=\columnwidth]{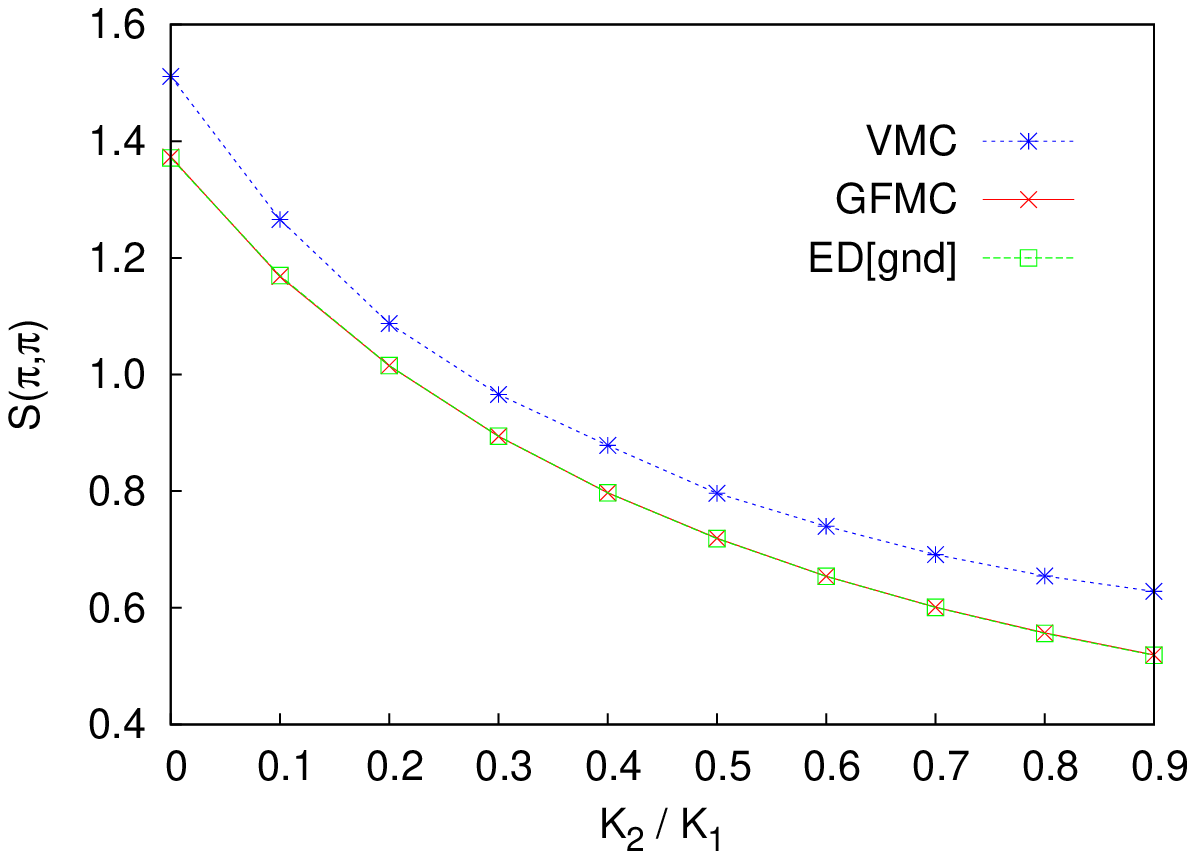}
  \caption{Top: Comparison of the 6$\times$6 lattice VMC and GFMC energies per site against the ED ground state and first excited state values for $0 \leq K_2 \leq 0.9$.  For $K_2 \geq 0.2$, the VMC energies are already closer to the ground state than to the first excited state;  however, despite such a good agreement in the energies, the VMC fails to identify a new phase for $K_2\geq 0.3$ as described in the text.
Bottom: Comparison of the corresponding VMC and GFMC structure factors $S(\pi,\pi)$ against the ED values.  The GFMC energies and structure factors show complete agreement with the ED values.}
  \label{fig:comparison_of_energies_SF_for_6x6_latt}
\end{figure}

In Fig.~\ref{fig:comparison_of_energies_SF_for_6x6_latt}, we test our GFMC setup against exact diagonalization (ED) calculations for the 6$\times$6 lattice.  All results are in the sector with 3 bosons in each row and in each column.  On both panels, the GFMC and ED results essentially coincide for the entire $K_2$ range shown.  Table~\ref{table:SF_6x6} summarizes the respective data for reference.  To check the accuracy of the trial wave functions, the VMC energies are also plotted in the top panel of Fig.~\ref{fig:comparison_of_energies_SF_for_6x6_latt} along with the first excited state ED energies.  For $K_2 \geq 0.2$, the VMC energies are already closer to the ground state than to the first excited state.  We caution the reader that despite this very good accuracy on the 6$\times$6 system, the VMC fails to identify another order that develops for larger $K_2$ and is found by the GFMC for larger sizes.

\begin{table}
  \begin{tabular*}{\columnwidth}{@{\extracolsep{\fill}}c|ccc|cccc}
    \hline
    $K_2$ & $E_{\rm VMC}$ & $E_{\rm GFMC}$ & $E_{\rm ED}$ & $S_{\rm VMC}^{(\pi,\pi)}$  & $S_{\rm GFMC}^{(\pi,\pi)}$ & $S_{\rm ED}^{(\pi,\pi)}$\\
    \hline\hline
    0.0 & -0.3714 & -0.39075 & -0.39075 & 1.511 & 1.368 & 1.371\\
    0.1 & -0.4078 & -0.42675 & -0.42675 & 1.266 & 1.168 & 1.170\\
    0.2 & -0.4467 & -0.46552 & -0.46552 & 1.087 & 1.015 & 1.015\\
    0.3 & -0.4875 & -0.50658 & -0.50658 & 0.966 & 0.894 & 0.894\\
    0.4 & -0.5303 & -0.54953 & -0.54953 & 0.878 & 0.797 & 0.797\\
    0.5 & -0.5744 & -0.59406 & -0.59406 & 0.796 & 0.719 & 0.719\\
    0.6 & -0.6206 & -0.63988 & -0.63988 & 0.740 & 0.654 & 0.654\\
    0.7 & -0.6672 & -0.68678 & -0.68678 & 0.691 & 0.601 & 0.601\\
    0.8 & -0.7139 & -0.73455 & -0.73456 & 0.655 & 0.557 & 0.556\\
    0.9 & -0.7614 & -0.78307 & -0.78307 & 0.628 & 0.519 & 0.518\\
    \hline
  \end{tabular*}
  \caption{ Comparison of the ground state energy and $S(\pi,\pi)$ obtained using the VMC, GFMC, and ED calculations for the 6$\times$6 lattice.  The energies are given in units of $K_1$ per lattice site. The GFMC results are essentially exact, and we treat them as such for larger sizes.}
  \label{table:SF_6x6}
\end{table}

\subsection{GFMC study of the $K_1$-$K_2$ model at half-filling}\label{sec:GFMC_half_filling}

\begin{figure}
\includegraphics[width=\columnwidth]{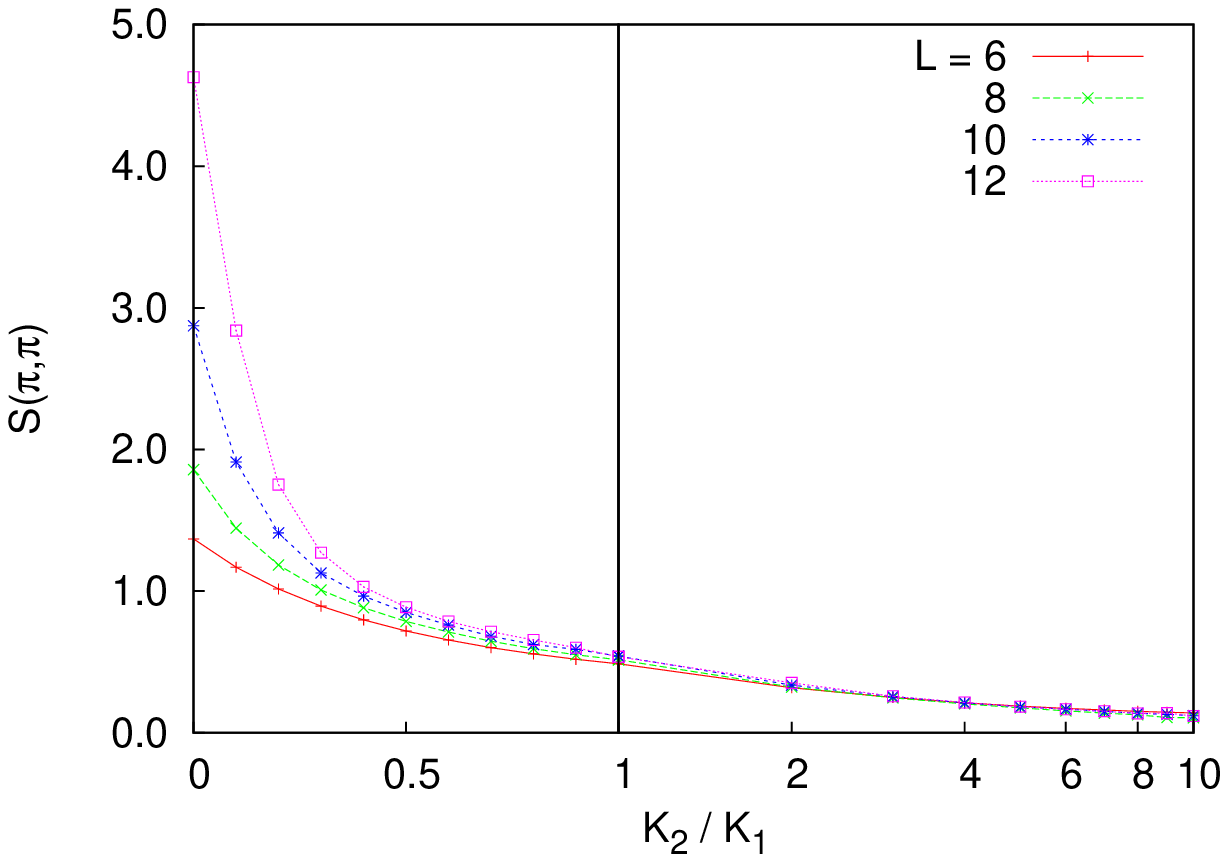}\\
\includegraphics[width=\columnwidth]{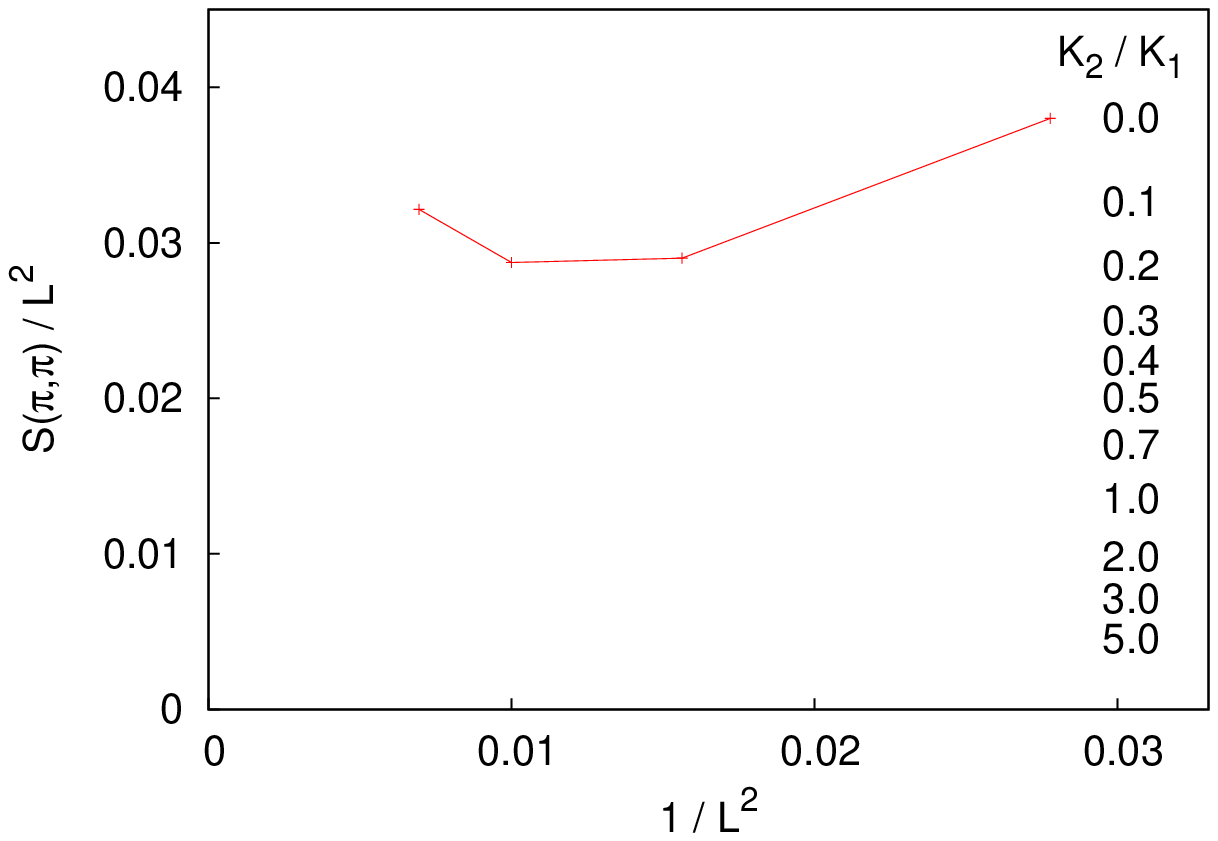}\\
\includegraphics[width=\columnwidth]{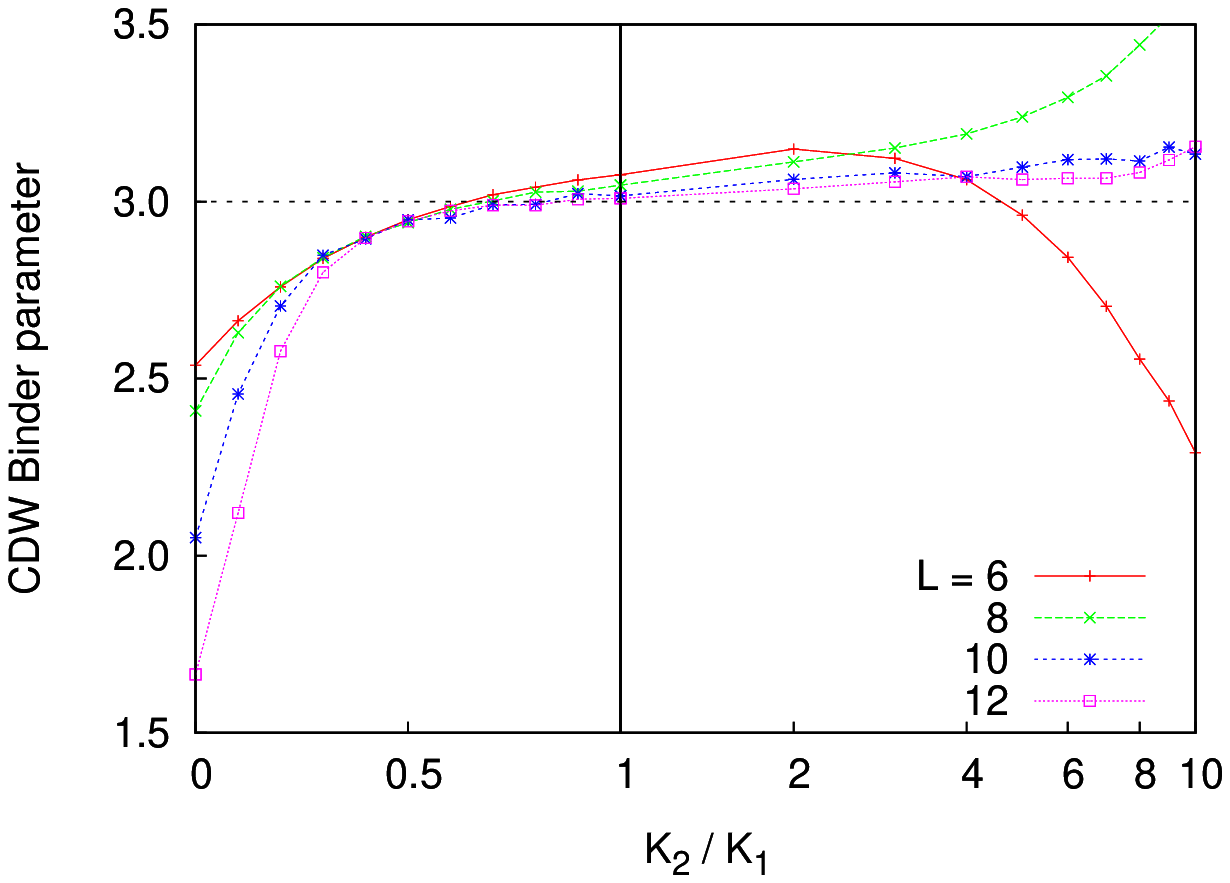}
\caption{
  Top: GFMC density structure factor $S(\pi,\pi)$ versus $K_2$ for periodic lattices with length $L=6,8,10$ and $12$.  Note that to show the data compactly, taken for $K_2=0$ to $1$ in steps of $0.1$ and $K_2=1$ to $10$ in steps of $1$, we used linear scale for the first range but log scale for the second range.  
  Middle: Finite size scaling of $S(\pi,\pi)$.
  Bottom: Binder ratio, Eq.~(\ref{Eq:binder}), for the CDW order parameter.  Note the apparently large finite size effect in the ratio, particularly for sizes $L=6$ and $8$ at large $K_2$, and also absence of clear Binder crossings.  Nevertheless, the Binder data is generally consistent with a lack of CDW order for $K_2 \gtrsim 0.3$, as can be seen from the ratio approaching the expected ``disordered'' value $3$ (shown with a dotted line).}
  \label{fig:Spipi}
\end{figure}

\begin{figure}
\includegraphics[width=\columnwidth]{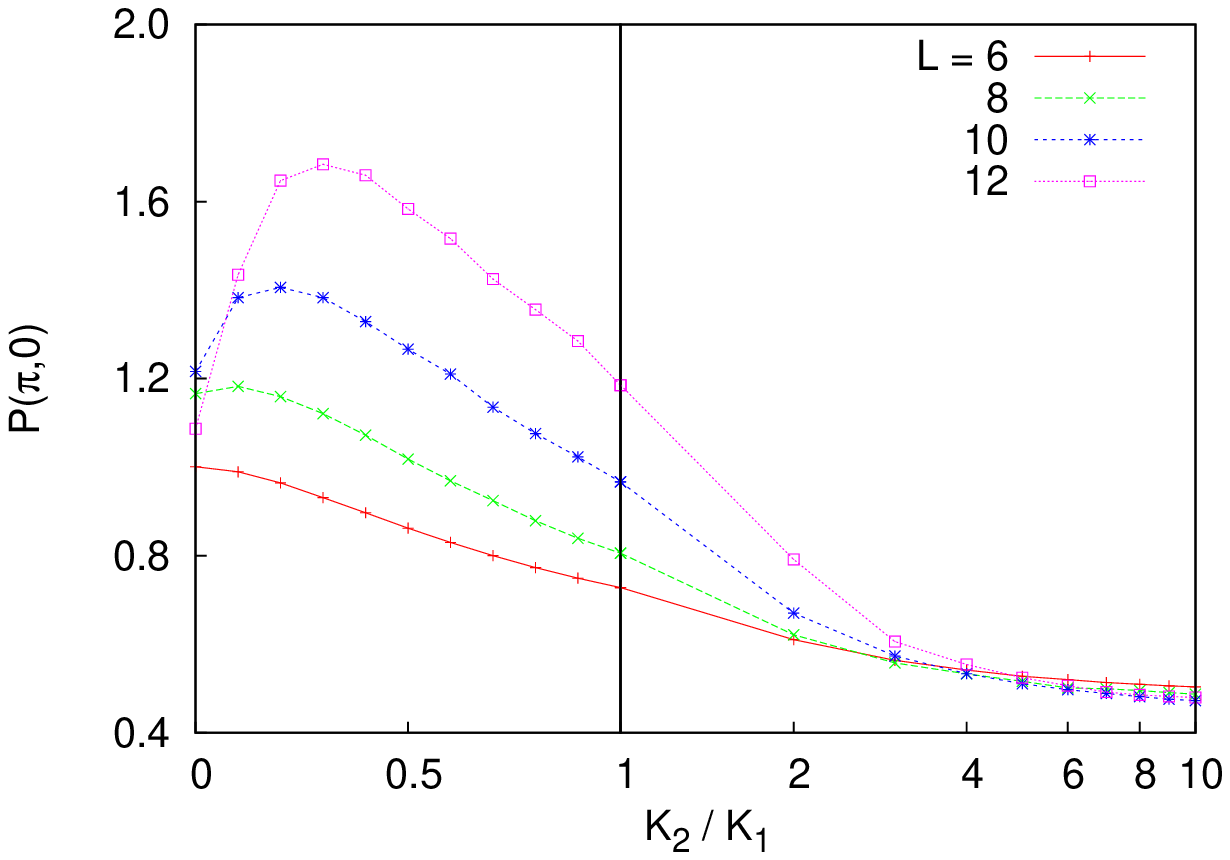}\\
\includegraphics[width=\columnwidth]{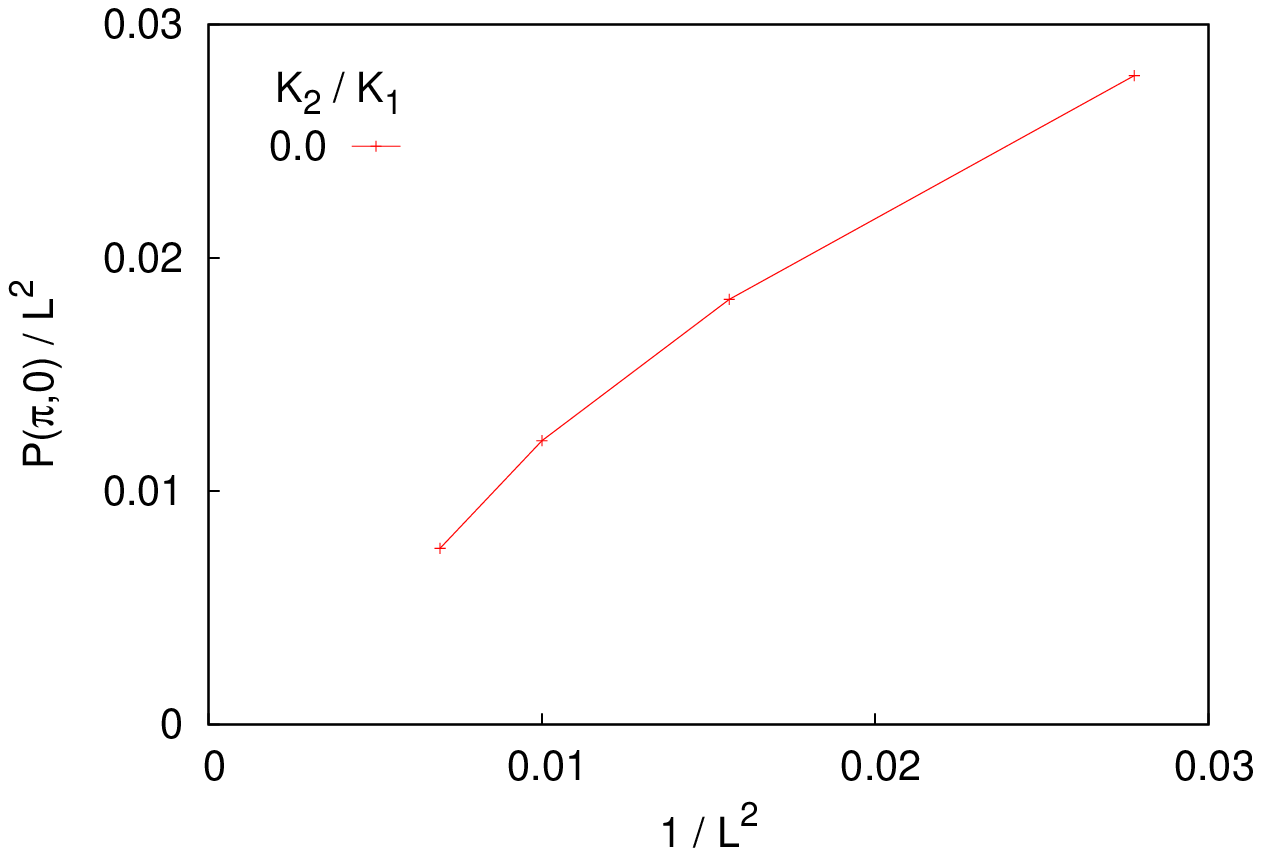}\\
\includegraphics[width=\columnwidth]{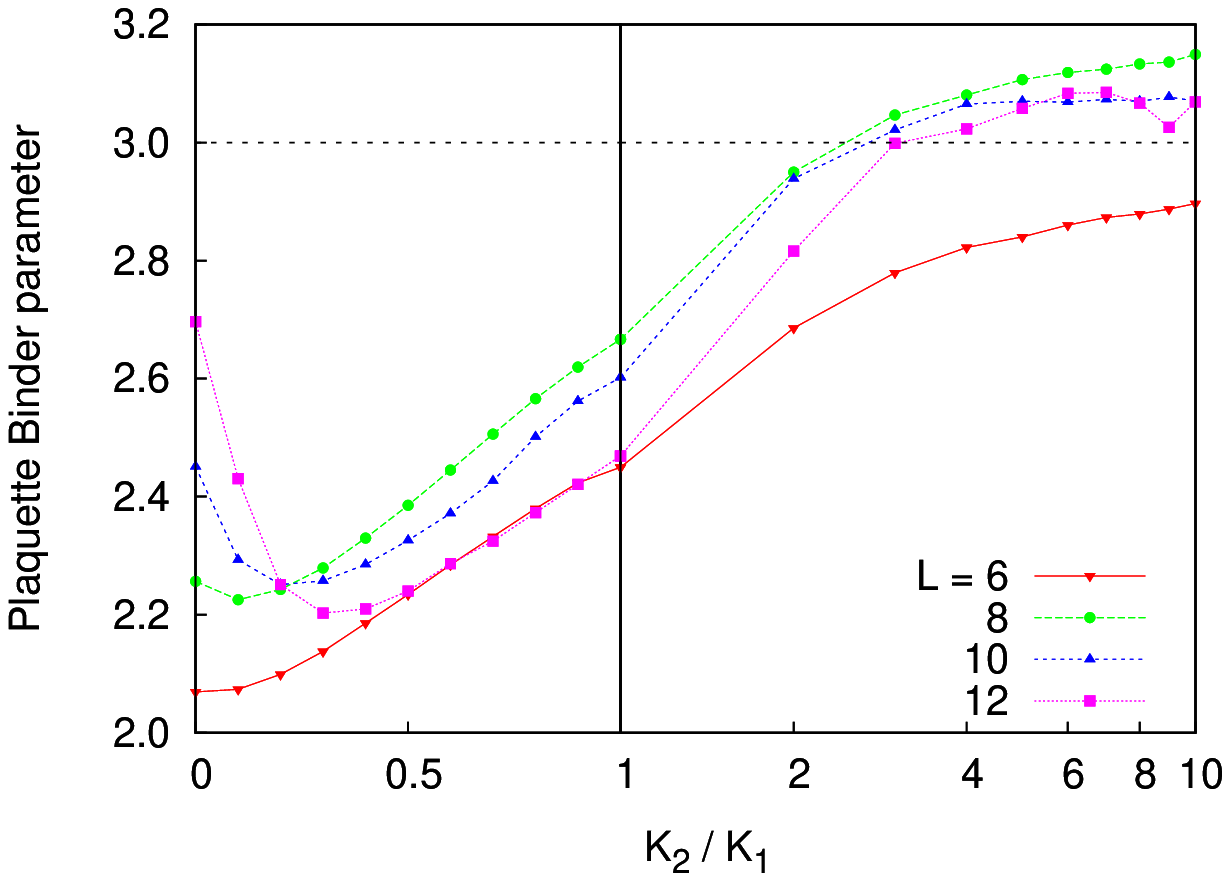}
\caption{
  Top: GFMC plaquette structure factor $P(\pi,0)$ versus $K_2$ for periodic lattices with length $L=6,8,10$ and $12$.  Linear scale is used for $K_2=0$ to $1$ in steps of $0.1$, and log scale for $K_2=1$ to $10$ in steps of $1$.
  Middle: Finite size scaling of $P(\pi,0)$.
  Bottom: Binder ratio, Eq.~(\ref{Eq:binder}), for the bond-solid order parameter.  The Binder data is consistent with no VBS order for small $K_2$ and VBS order for intermediate $K_2$.  The observation of Binder ratios exceeding the disordered value 3 suggests that there is no order for large $K_2$, although there is no clear crossing going to the large $K_2$ phase (note that similar somewhat unusual behavior is also seen in the CDW Binder data when the CDW order disappeared.)
}
  \label{fig:Ppi0}
\end{figure}

We now proceed to apply this numerical tool to characterize the ground states of the $K_1$-$K_2$ ring model.  The top panels in Figs.~\ref{fig:Spipi} and \ref{fig:Ppi0} show the density structure factor $S(\pi,\pi)$ and the plaquette structure factor $P(\pi,0)$ plotted against $K_2$ for lattice sizes ranging from $L=6$ to $12$.  Between $K_2=0$ and $0.4$, $S(\pi,\pi)$ increases strongly with $L$ while the size dependence weakens with $K_2$.  This coincides with a strengthening size dependence of $P(\pi,0)$.  Beyond $K_2=0.4$, the charge order is absent while the plaquette order now dominates in the range up to $K_2\approx 4$.  For still larger $K_2$, $P(\pi,0)$ becomes very weakly dependent on lattice size.
We do not observe any other strong feature in $S(q_x,q_y)$ and $P(q_x,q_y)$ over the full Brillouin zone.  Thus we identify the ($\pi$,$\pi$) CDW for $0 \leq K_2 < 0.4$, a $(\pi,0)$ bond-solid--type phase for the intermediate $K_2$ region, and tentatively an EBL phase for $K_2 > 4$.   The middle panels in Figs.~\ref{fig:Spipi} and \ref{fig:Ppi0} show finite-size scalings of the respective order parameters which support these conclusions.

The bottom panels in Figs.~\ref{fig:Spipi} and \ref{fig:Ppi0} show the corresponding Binder ratios defined as
\begin{equation}
  \text{Binder ratio} = \frac{\la |M|^4 \ra}{\la |M|^2\ra^2}
  \label{Eq:binder}
\end{equation}
with $M_{\rm CDW} = \sum_{\bf r} e^{i (\pi,\pi) \cdot {\bf r}} n_{\bf r}$ or $M_{\rm VBS} = \sum_{\bf r} e^{i (\pi,0) \cdot {\bf r}} (P^{11}_{\bf r})^2$ (so $\la |M|^2\ra$ are simply proportional to the already discussed structure factors).  The Binder ratios provide additional information about the fluctuations of the order parameters (via the measurement $\la |M|^4\ra$) and are expected to approach $1$ in the presence of the order and $3$ in the absence of the order.  Such change in the behavior is clearly seen when the CDW order disappears and the plaquette order appears near $K_2 \sim 0.3 - 0.4$.  Note, however, that the familiar Binder crossing technique apparently does not work for the CDW order parameter for our sizes, even though we are confident that the CDW order disappears (also supported by the values $\approx 3$ of the Binder ratio itself).  Note also strong and non-systematic size dependence, particularly for the smallest $L=6$.
In the plaquette Binder ratio, we see lack of order for small $K_2$, appearance of order for intermediate $K_2$, and apparently ``disordered'' Binder values for $K_2 \gtrsim 4$ (which is consistent with the absence of the plaquette order), but no clear crossings for this transition.  Although the Binder data does not clearly give us the critical value of $K_2$ for the transition to the disordered phase, its limiting value strongly suggests that there is no $(\pi,0)$ or $(0,\pi)$ plaquette order at large $K_2$.  An additional lesson from this study is that we should be aware of particular strong finite size effects in this system.

Our identification of the VBS order for the intermediate $K_2$ region is further helped by measurement of the bond structure factor $B_x(q_x,q_y)$ defined in Eq.~(\ref{Eq:Bqxqy}).  For this measurement (not shown), similarly to the plaquette structure factor, we observe $(\pi,0)$ and $(0,\pi)$ Bragg peaks but no peak at $(\pi,\pi)$.  This is more consistent with a ``columnar'' VBS order rather than a plaquette order, and is also similar to the phase found by Sandvik\etal\cite{Sandvik2002} in the $J$-$K$ model for $8 \lesssim K/J \lesssim 14$.  Our finding of the same VBS state may in fact be related, since $J$ added to the pure $K_1$ model may induce effective $K_2$ ring exchanges frustrating the CDW while still remaining in the Mott insulator.

\begin{figure*}
  \center
  \includegraphics[trim=20 0 0 0, scale=0.67]{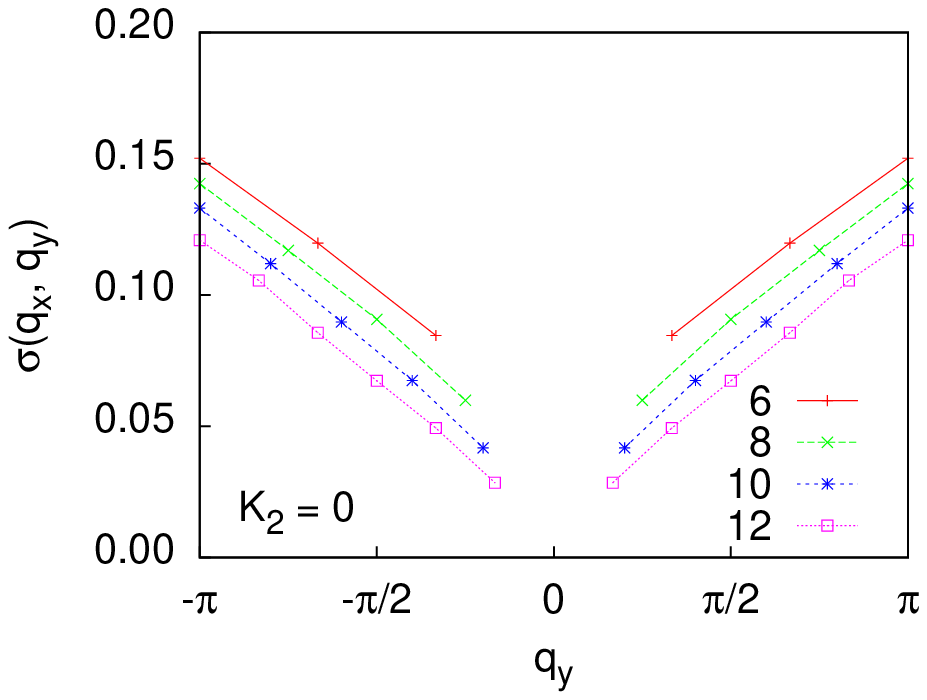}  \includegraphics[trim=50 0 0 0, scale=0.67]{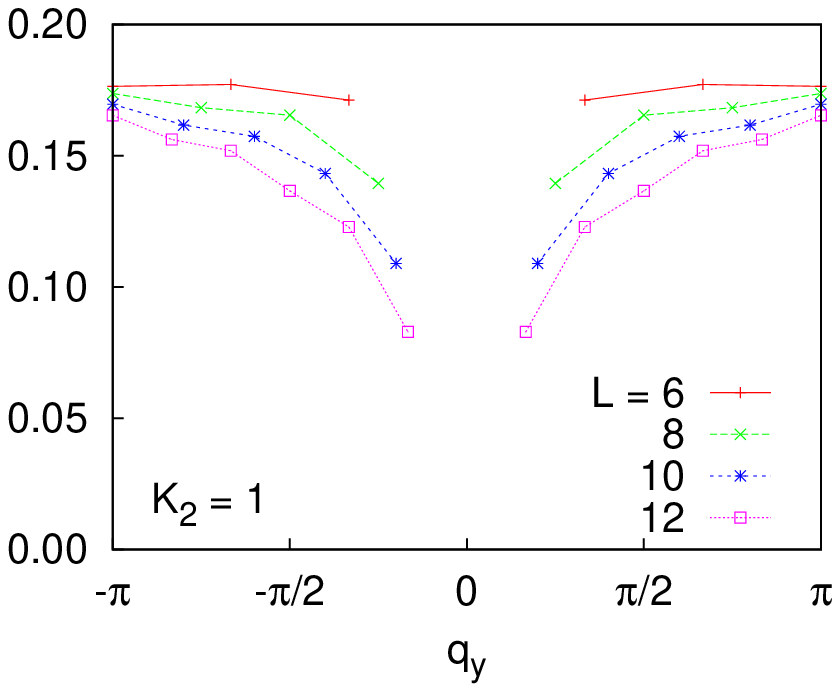} \includegraphics[trim=40 0 20 0, scale=0.67]{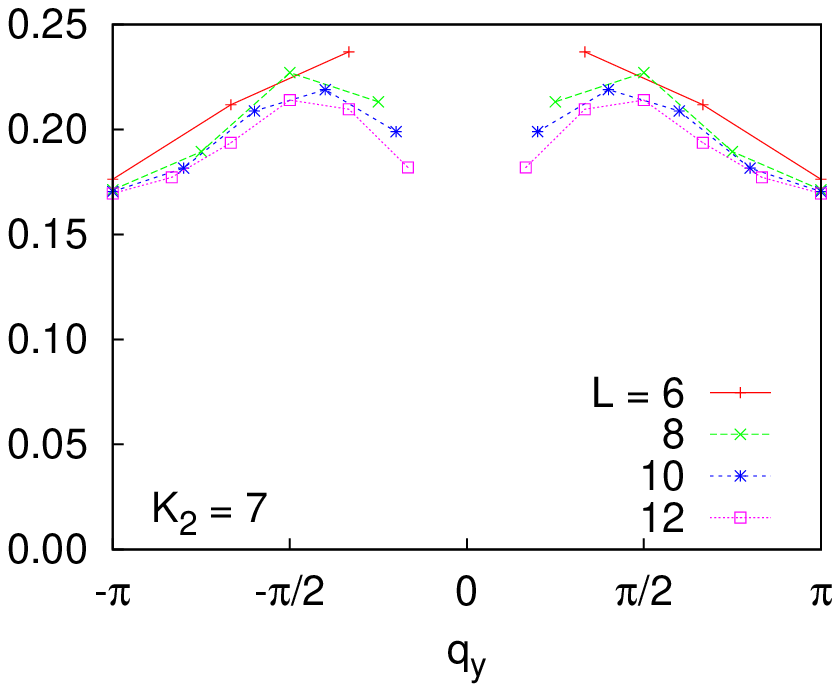}
  \caption{``Cross analysis'' of the density structure factors, plotting ``normalized slopes'' $\sigma(q_x=2\pi/L, q_y)$, Eq.~(\ref{sigma}), versus $q_y$ for $L=6$ to $12$.
Left: $K_2 = 0$ in the CDW phase; the vanishing of the slopes is consistent with nonsingular behavior expected in an incompressible system.
Middle: $K_2 = 1$ in the VBS phase; the vanishing of the slopes continues to be present and can be used as an indication of the EBL instability even if we did not know the resulting order.
Right: $K_2 = 7$ (note different vertical scale); the data shows non-zero normalized slopes and only weak renormalizations, therefore suggesting a stable EBL phase.
}
  \label{fig:GFMC_wavefunction_transition}
\end{figure*}

To get a more complete picture, we examine the long wavelength behavior of the density structure factor using the ``cross analysis'' of Sec.~\ref{sec:VMC_formal}.   Figure \ref{fig:GFMC_wavefunction_transition} shows the ``normalized slopes'' $\sigma(q_x \!=\! 2\pi/L, q_y)$, Eq.~(\ref{sigma}), for $L=6$ to 12.  The left panel shows the results for $K_2=0$, 
which we already know is in the CDW phase from the presence of the ($\pi$,$\pi$) Bragg peak.   We clearly observe a Mott-like incompressible behavior where the slopes vanish.  This is similar to the earlier formal wave function study with the CDW.  The Mott-like dependence of $S(q_x,q_y)$ at long wavelengths continues to be present after the charge order disappears for $K_2 \gtrsim 0.4$.  This is illustrated in the middle panel for $K_2=1$, at which the bond-solid ordering is already established in Fig.~\ref{fig:Ppi0}.  Our ``cross analysis'' therefore provides an independent detection of the instability to a different solid.

In the right panel of Fig.~\ref{fig:GFMC_wavefunction_transition} for $K_2=7$,  our small-lattice data appears to suggest that $S(q_x,q_y)$ does have the V-shaped singularity along the lines $q_x=0$ or $q_y=0$.  This would mean that the bond-solid ordering exists only at intermediate $K_2$ and hence, possibly realizing the EBL phase at large $K_2$.   However, this panel also reveals a weak downwards renormalization of the slopes upon increasing $L$ and we therefore do not rule out the possibility of the EBL behavior disappearing at much larger lattice sizes.

\begin{figure}
  \centering
  \includegraphics[trim=15 20 10 60,width=\columnwidth]{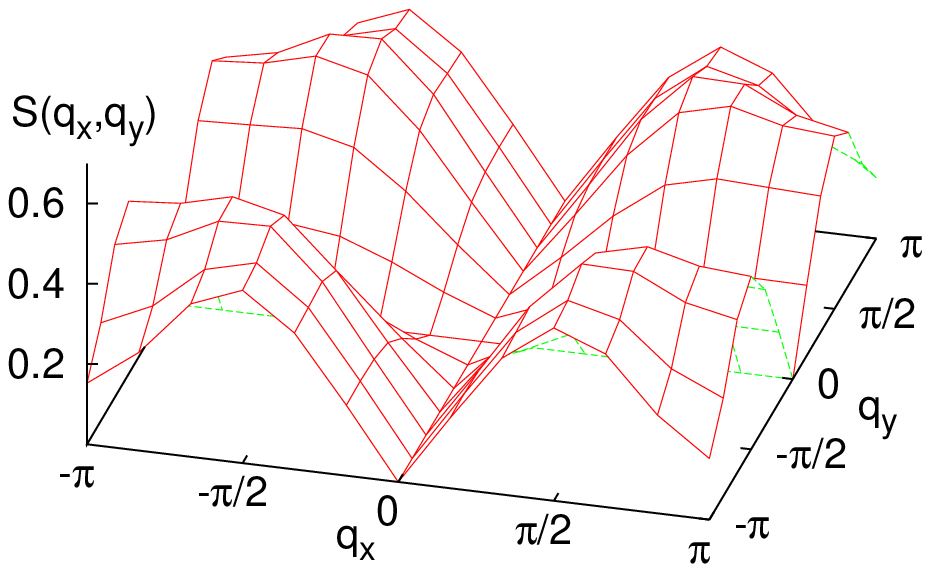}\\
  \includegraphics[trim=00 10 10 55,width=\columnwidth]{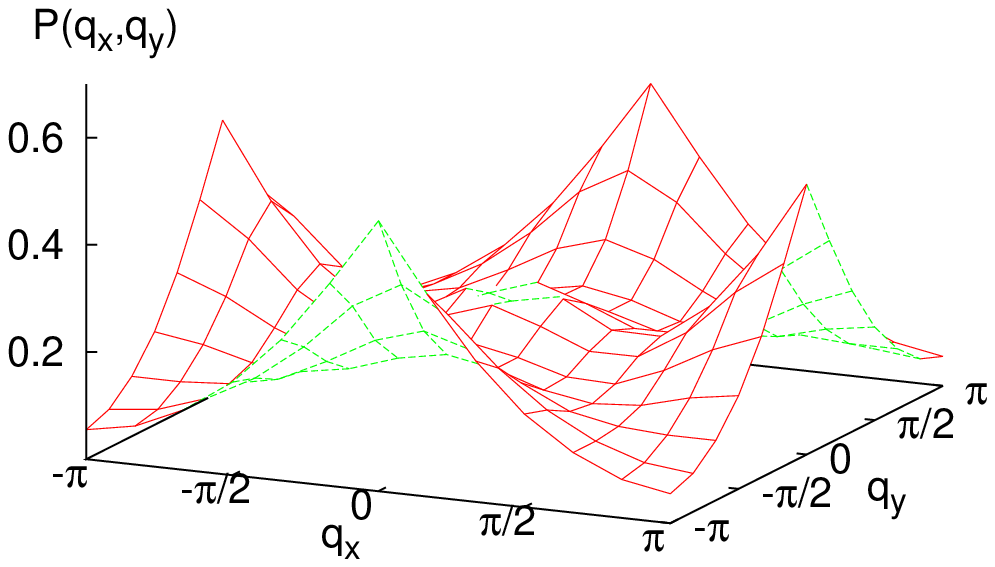}\\
  \includegraphics[trim=00 10 10 3, scale=0.75]{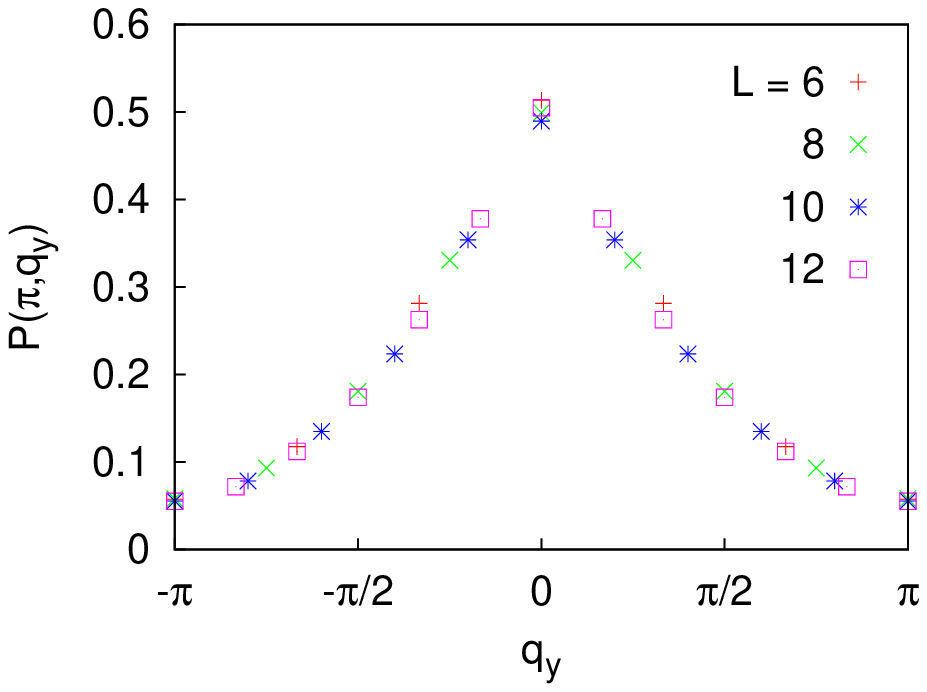}
  \caption{Top: GFMC density structure factor $S(q_x, q_y)$ for a 12$\times$12 lattice at $K_2=7$.  Middle: GFMC plaquette structure factor $P(q_x, q_y)$ for the same system (${\bf q} = {\bf 0}$ point not calculated). Bottom: Cut of $P(q_x, q_y)$ at $q_x=\pi$ for $L=6$ to $12$.}
  \label{fig:SF_12x12_k5}
\end{figure}

A rigorous confrontation of the large $K_2$ region requires a study on much larger lattices, but this is beyond the capability of our present numerical setup.  We instead examine the structure factors $S(q_x,q_y)$ and $P(q_x,q_y)$ over the entire Brillouin zone and look for signatures of the EBL phase as well as potential instabilities.  The top panel in Fig.~\ref{fig:SF_12x12_k5} shows the density structure factor at $K_2=7$ and clear absence of any CDW ordering.  Here, we highlight the presence of the long wavelength EBL signature near the lines $q_x=0$ and $q_y=0$ (this characteristic cross has already been analyzed in Fig.~\ref{fig:GFMC_wavefunction_transition}).  The middle panel in Fig.~\ref{fig:SF_12x12_k5} shows the plaquette structure factor for the same system, which again does not show bond or plaquette ordering.  Despite the potential instability hinted by the $P(\pi,0)$ and $P(0,\pi)$ cusps, the size independence of the plaquette structure factor along the cut $q_x=\pi$ shown in the bottom panel of Fig.~\ref{fig:SF_12x12_k5} gives us some confidence that the EBL phase may indeed be realized in the large $K_2$ regime at half-filling.

\begin{figure}
  \center
  \includegraphics[width=\columnwidth]{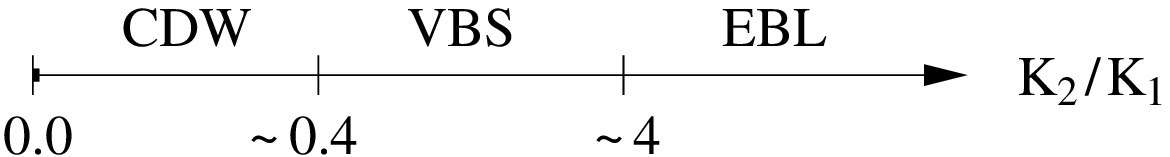}
  \caption{GFMC phase diagram for the $K_1$-$K_2$ model on the half-filled square lattice.  We found the $(\pi, \pi)$ CDW for $K_2\lesssim 0.4$, the $(\pi, 0)$ VBS for intermediate values of $K_2$, and possibly the EBL phase for $K_2\gtrsim 4$.}
  \label{fig:GFMC_phase_diagram}
\end{figure}

Figure~\ref{fig:GFMC_phase_diagram} summarizes the unbiased GFMC phase diagram obtained for the $K_1$-$K_2$ model at half-filling.

\subsection{More detailed comparison with the EBL theory and interpretations}

We now discuss how the presented results at half-filling may fit into the EBL theory framework reviewed in Appendix~\ref{app:EBLprecis}.  The EBL is characterized by an ``EBL phase stiffness.''  For example, in the spin wave theory Eq.~(\ref{eq:hamiltonian_SW}), the EBL stiffness is simply $K/U$, which we will parametrize by $\kappa = \sqrt{K/U}$.  More generally, the EBL stiffness is a {\it function} on the cross lines $q_x = 0$ or $q_y = 0$, with convenient parametrization by $\kappa(q_y) = \sqrt{\cK(0, q_y) / \cU(0, q_y)}$ in the notation from Appendix~\ref{app:EBLprecis}.  As described there, $\kappa(q_y)$ determines exponents in various power-law correlations along lattice directions, which in turn determine stability of the EBL.

We do not have a direct access to the EBL phase stiffness in our setup, but we can crudely monitor its behavior using the characteristic cross in the density structure factor.  Specifically, from the EBL theory result Eq.~(\ref{cross}), we have for the ``normalized slopes'' Eq.~(\ref{sigma}):
\begin{equation}
\sigma(0, q_y) = \frac{\kappa(q_y)}{2} \times |C_\rho(0, q_y)|^2~.
\label{sigma2kappa}
\end{equation}
Here $C_\rho({\bf q})$ is a non-universal function of order 1.  In the spin-wave theory, $C_\rho({\bf q}) = 1$.  In general, $C_\rho({\bf q} \to {\bf 0}) = 1$, and we conjecture also that $C_\rho(q_x \to 0, q_y) = 1$ for any $q_y$, but we do not know for sure.  If we can indeed ignore the $|C_\rho(0, q_y)|^2$ factor in Eq.~(\ref{sigma2kappa}), we can then view our ``cross analysis'' presented earlier as a crude measure of the effective EBL stiffness on the corresponding length scales.  When the measured $\sigma(q_{\rm min} \!=\! 2\pi/L, q_y)$ renormalizes down to small values upon increasing the system size, the EBL is unstable; when $\sigma(2\pi/L, q_y)$ stays finite, the EBL is stable.

The theoretical stability of the EBL requires the stiffness to be sufficiently large, and the condition is particularly stringent at half-filling because of allowed Umklapp interactions.  The corresponding scaling dimensions are given in terms of $\kappa(q_y)$ in Appendix~\ref{app:EBLprecis}.  For a $q_y$-independent $\kappa$, the leading Umklapp has scaling dimension given by Eq.~(\ref{u1pwr}) and is irrelevant if
\begin{equation}
\kappa > 3/8 ~,
\label{u1irrel}
\end{equation}
which we can take as a rough guide at half-filling.  Our EBL cross analysis of the density structure factor gives $\sigma \approx \kappa/2$, so to establish stability we would like to see $\sigma \geq 3/16 = 0.1875$.  For $K_2 = 7$ this is satisfied on average for sizes $L=6$ to $12$, but the larger sizes come close to the threshold.  The state may be somewhat more stable in that the measured $\sigma(q_y)$ is largest near $q_y=\pi/2$ and the particular ``average'' of $\kappa(q_y)$ that one needs has the main weight in the middle of the region $[0, \pi]$, cf.\ Eq.~(\ref{uj}).  [We also want to repeat that we are not sure whether we can ignore the factor $|C_\rho|^2$ in Eq.~(\ref{sigma2kappa}) and unambiguously extract $\kappa$ from $\sigma$.]
From the absence of any orders, we conjecture that this point is stable; of course, if the instability is very weak, we may be not detecting the order on our length scales.  

For $K_2 = 0$ and $K_2 = 1$, the effective EBL stiffness parameters as estimated by the cross analysis in Fig.~\ref{fig:GFMC_wavefunction_transition} are below critical or border-line critical already for the size $L=6$ and then quickly renormalize down upon increasing $L$, consistent with our finding of the instability of the EBL towards boson solid phases.

As reviewed in more technical terms in Appendix~\ref{app:EBLprecis}, when the EBL is unstable at half-filling, the natural outcomes are a $(\pi,\pi)$ CDW or a plaquette solid with period 2 in both lattice directions.  The outcome depends on the sign of some effective couplings.  If an effective nearest neighbor repulsion dominates, the solid locks into the CDW, while if the second-neighbor repulsion dominates, the solid locks into the plaquette state.  We then propose that, as we increase the $K_2$, while the EBL remains unstable (e.g., as detected by the cross analysis), at some point the sign of the locking switches from the CDW to the bond-solid.  (At present, we do not know how to realize the columnar VBS out of the EBL theory, but usually columnar and plaquette orders are related\cite{ReadSachdev1989, Lannert2001, Senthil2004science, Senthil2004} and perhaps we are missing some physics ingredients in the theory that would enable the columnar VBS.)  As we further increase $K_2$, we conjecture that the Umklapp eventually becomes irrelevant and the stable EBL is realized.  In this scenario, we do not anticipate any other instability, so if the presented large $K_2$ region is eventually unstable, the simplest possibility is that it will have a very small VBS order.

\subsubsection{Interpretation of the failure of the VMC at $\rho=1/2$}
\label{subsec:VMCfailure}
In light of the above stability considerations, we now briefly discuss the failure of the formal wave function study in Sec.~\ref{sec:VMC} to detect the EBL instability in the intermediate $K_2$ regime.  We presented mainly the one-parameter wave function that can capture only the EBL or CDW.  However, we also considered a two-parameter wave function where in the spin wave theory like Eq.~(\ref{eq:hamiltonian_SW}) we include both $K_1$ and $K_2$ ring terms (in fact, this was used throughout to obtain improved initial states for the GFMC projection).  We found that such a wave function, depending on the parameters, can realize also the VBS state on the same footing as the CDW.\cite{unpublished}  Nevertheless, in the energetics study in the intermediate $K_2$ regime, the optimized two-parameter wave function produces a liquid.

The one-parameter example from Sec.~\ref{sec:VMC} is sufficient for our discussion.  The wave function is constructed from the EBL spin wave theory and it seems reasonable to take the EBL parameter as $\kappa = 1/W$.  Consider now $W=4.2$ shown in the middle panel in Fig.~\ref{fig:VMC_wavefunction_transition}, where the normalized slopes in the density structure factor approached the expected value $\sigma_{\rm SW} = 1/(2W)$ and where we concluded that the VMC wave function is in the liquid phase.  However, such stiffness $\kappa = 0.24$ strongly violates the stability condition Eq.~(\ref{u1irrel}).  Therefore, we appear to have a situation where the formal wave function is a liquid with a stiffness that is too small for the full EBL theory to be stable.  

This is reminiscent of what happens when one formally considers a Jastrow-type wave function for 1D hard-core bosons,
$\Psi_{1D} = \Pi_{i<j} |\sin[\pi(x_i - x_j)/L]|^\nu$.
The wave function describes a Luttinger liquid of bosons with the Luttinger parameter $g = 1/\nu$.  On a half-filled chain, the Luttinger liquid becomes unstable to a staggered CDW when $g < 1/2$ corresponding to $\nu > 2$.  However, as discussed in Ref.~\onlinecite{Narayan1999}, the above wave function remains liquid until $\nu$ exceeds 4, and only then the CDW order develops.  Thus, in this 1D example, the condition for the formal stability of the wave function is different from that in the full theory.

Assuming similar phenomenon for the formal EBL wave function, we can then speculate what happened in our variational study.  The wave function parameter is found by optimizing the energetics and it roughly captures the bare EBL stiffness on the scale of few lattice spacings.  For small $K_2$, this is already in the regime where both the wave function and the full EBL theory are unstable.  However, for intermediate $K_2$ the optimized parameters happen to be in the range where the wave function is stable while the full EBL theory is not, hence the failure of our VMC.  Of course, once we suspect boson-solid phases, the use of the few-parameter wave functions motivated from the liquid side becomes inadequate.  In the variational approach, more parameters also allowing the Jastrow pseudopotentials to become more long-ranged would be needed,\cite{Capello2007short, Capello2008long} while in the present study the correct physics is brought by the GFMC projection.

\section{Study of the $K_1$-$K_2$ model for $\rho<1/2$}\label{sec:general_densities}

When we step away from half-filling, the Umklapp terms discussed above are no longer allowed.  While the EBL may still be unstable due to non-Umklapps, they are typically less relevant (see Appendix~\ref{app:EBLprecis}).  However, here one also competes against phase separation at low densities.  Previous studies\cite{Melko2004, Rousseau2004, Rousseau2005, Motrunich2007} of boson models with 1$\times$1 ring exchanges found that ring interactions induce strong tendency to phase separation, since they are operative only when bosons are nearby.  The more extended $K_2$ ring interactions can somewhat offset this tendency and produce a stable uniform EBL regime over a wider range of densities below half-filling.

We would like to point out that our restricted Hilbert space with equal boson number in each row and each column does not preclude phase separation.  For example, basis states with preferential clumping along a diagonal or in blocks along the diagonal are present in our Hilbert space.  In fact, we observe regimes of phase separation in the VMC and GFMC simulations which will be discussed below.  We detect the phase separation in the Monte Carlo simulations either by monitoring snapshots of the real-space boson configurations, or by looking at the structure factors in momentum space, where it is revealed by the presence of strong peaks at the smallest wavevectors. To further check the results, we start our simulations from both uniform (random) boson configurations and from half-filled diagonal stripes.  We verified that, independent of the initial configurations, our simulations converge to uniform states for densities close to 1/2 but phase-separate for low densities.

It may be true that in our working sector the phase separation is somewhat suppressed in finite samples, since other shapes of clumped regions are not allowed.  However, this effect should decrease with increasing system size, and it is likely that our sizes already crudely capture such local energetics tendencies as to whether the system wants to stay uniform or phase-separate.

Below, we present results of a VMC energetics study on a 24$\times$24 lattice and results of a GFMC study on lattices with $L\leq 12$.  We find that a stable EBL phase is present in a window $\rho\in (0.4,0.5)$ for $K_2$ as small as $0.5$.

\subsection{VMC results for 24$\times$24 lattice}\label{sec:VMC_extended}

\begin{figure}
  \centering
  \includegraphics[width=\columnwidth]{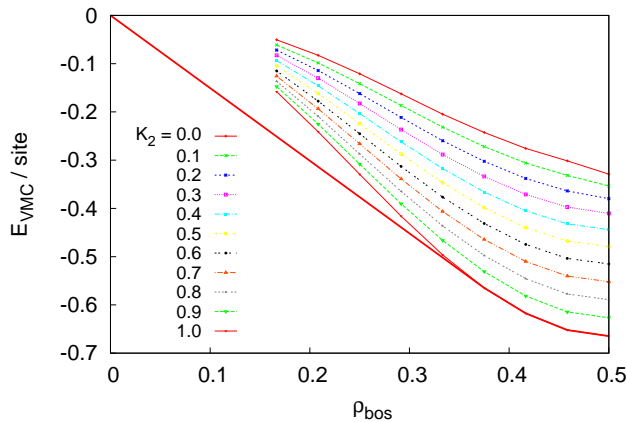}
  \caption{Trial energies of the optimized EBL wave function measured on a 24$\times$24 lattice. Using Maxwell construction (illustrated for $K_2=1$), the critical density dividing the uniform liquid regime and the phase-separated regime can be determined for each $K_2$. We found a decrease in the critical density from $0.45$ to $0.37$ as $K_2$ increases from $0$ to $1$.}
  \label{fig:vmc_E_rho}
\end{figure}

For our VMC energetics study of the uniform liquid phase on the 24$\times$24 lattice, we use the single-parameter EBL wave function from Eq.~(\ref{eq:PsiEBL}).  We consider total boson numbers in multiples of 24, as appropriate for the Hilbert space with equal number of bosons in each row and column.  Figure~\ref{fig:vmc_E_rho} shows the trial energy per site for the optimized EBL wave function plotted against boson density.  Away from half-filling, uniform VMC liquids are obtained for all data points shown in the figure.   For $\rho\lesssim 0.13$, we already detect phase separation in the VMC simulations (data points for these cases were excluded from the plot).  Among the uniform liquids shown on the plot, not every point produces a stable liquid.   We can use Maxwell construction (illustrated for $K_2=1.0$) to determine the critical density $\rho_c$ dividing the stable uniform liquid regime ($\rho_c < \rho < 1/2$) and the phase-separated regime ($\rho < \rho_c$), the latter comprising a uniform liquid region and an empty region on the lattice.  We found a decrease in the critical density $\rho_c$ from approximately $0.45$ to $0.37$ as $K_2$ increases from $0$ to $1.0$.  Thus, the additional $K_2$ ring terms indeed help to widen the stable regime of the uniform EBL phase in the $K_1$-$K_2$ model.

We should of course be cautious taking the VMC results too literally, given the described experience with the failures of the VMC at half-filling.  However, we are probably in a better position here in that the ``bare EBL stiffnesses'' are such that both the wave function and the full theory are stable.  Specifically, for $K_2 \geq 0.2$ and all densities $\rho < 1/2$, the optimal $W$ is smaller than $3.5$ and is further decreasing with increasing $K_2$.  In the absence of Umklapps, the most important non-Umklapp has scaling dimension $8 \sqrt{K/U} = 8/W > 2$ (see Appendix~\ref{subapp:gendens}); hence, all residual interactions are irrelevant and the full EBL theory is stable.  While the VMC results are suggestive, the ultimate determination of the phase diagram requires unbiased approaches.

\subsection{GFMC results for $L\leq 12$ lattices}\label{sec:GFMC_extended}

\begin{figure}
  \centering
  \includegraphics[width=\columnwidth]{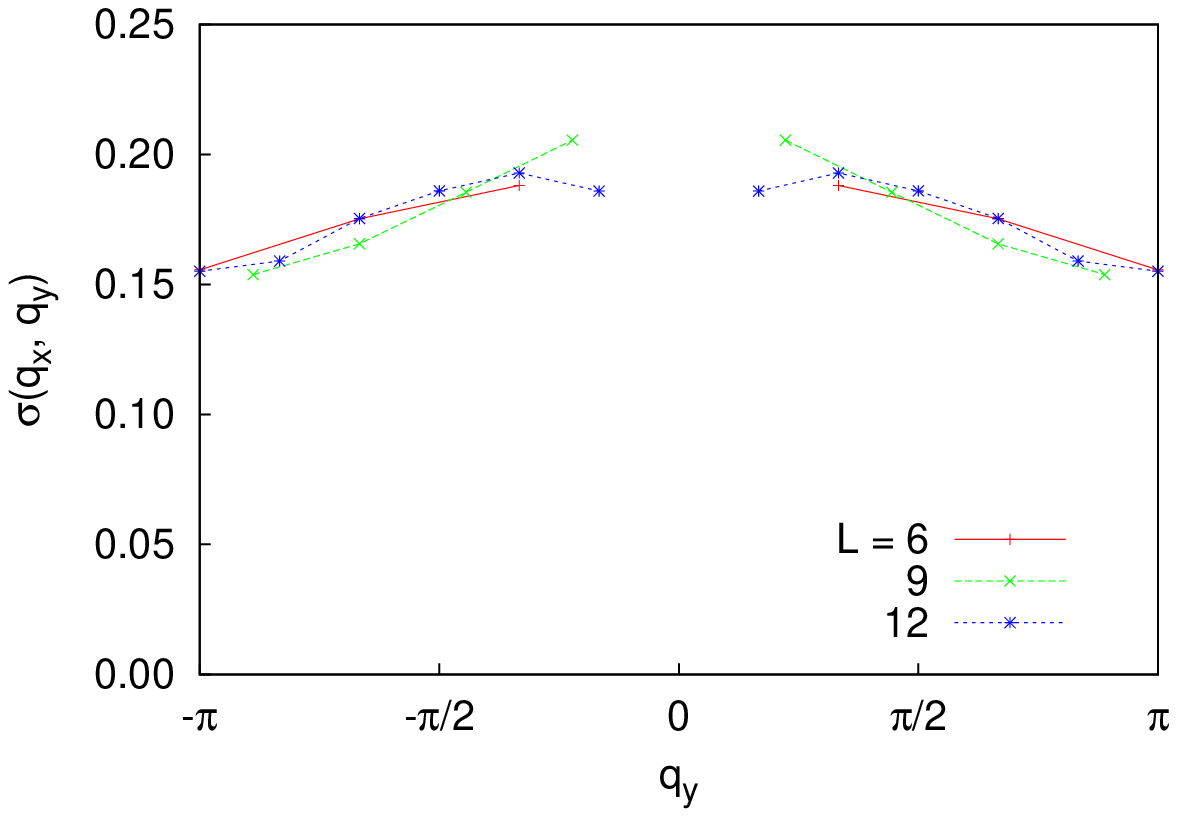}\\
  \includegraphics[trim=0 0 30 20,width=\columnwidth]{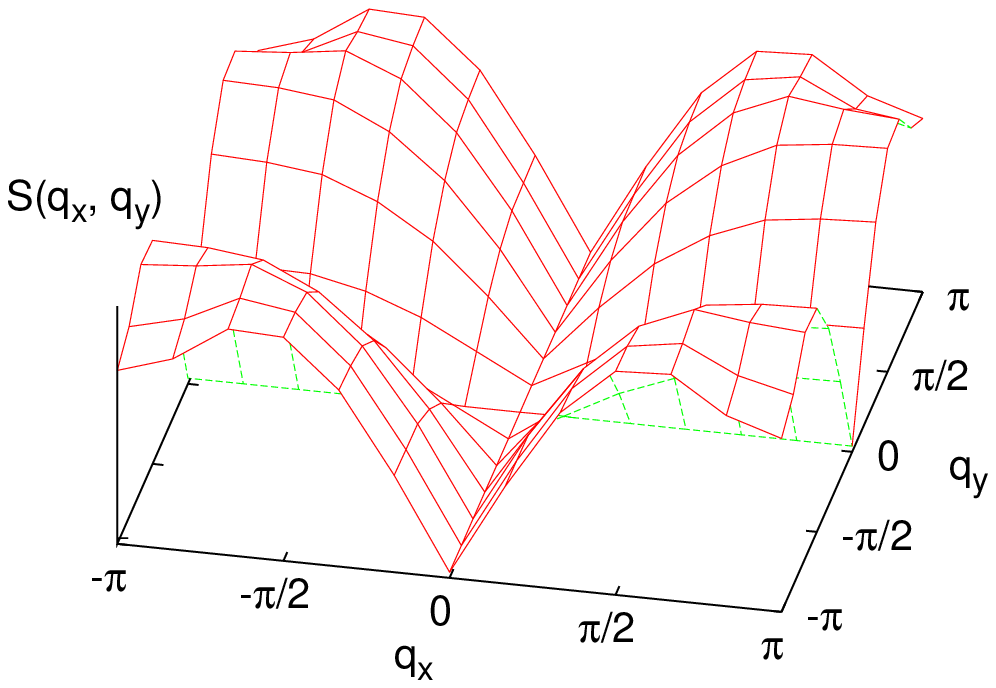}
  \caption{The GFMC density structure factor $S(q_x, q_y)$ on the 12$\times$12 lattice at density $\rho=1/3$ and $K_2=1.0$.  Top: ``Cross analysis'' showing normalized slopes $\sigma(q_x \!=\! 2\pi/L, q_y)$, Eq.~(\ref{sigma}), versus $q_y$.    Bottom: $S(q_x, q_y)$ in the full Brillouin zone.  Besides the ``cross'' signature, notice also (weak) ``$2k_F$ ridges'' [here lines $(\pm 2\pi/3, q_y)$ and $(q_x, \pm 2\pi/3)$] which are typically present for all the uniform liquids observed at density less than half.}
  \label{fig:GFMC_rho_cross}
\end{figure}

Next, we discuss the results of our GFMC study on the 12$\times$12 lattice for two selected points in parameter space, $K_2=0.5$ and $1.0$.   We consider densities varying from 2 up to 5 bosons per row and per column (corresponding to densities $\rho = 1/6$ up to $5/12$).  For 2 and 3 bosons per row at $K_2 = 0.5$ and 2 bosons per row at $K_2 = 1.0$, we already see signatures of phase separation both in real space and momentum space.   For higher boson number per row and per column, we do not see any sign of the phase separation. This allows us to conclude that the 12$\times$12 system is in a stable uniform phase (i.e. without phase separation) for $1/3 \lesssim \rho < 1/2$.

To determine whether the uniform phase realizes the EBL, we examine the GFMC density and plaquette structure factors for any sign of instability to CDW or bond-solid ordering, but we do not observe any strong peak in $S(q_x,q_y)$ or $P(q_x,q_y)$.   We apply the ``cross analysis'' of earlier sections to study the long wavelength behavior of the density structure factor.   The top panel of Fig.~\ref{fig:GFMC_rho_cross} shows the analysis done for $K_2=1.0$ and $\rho=1/3$, which is carried out on $L\times L$ lattices with $L=6$, 9 and 12.   Note that the crude stability condition at generic densities is $\sigma \approx \kappa/2 = (1/2) \sqrt{K/U} > 1/8$ and is safely satisfied.  Note also that the bare EBL stiffness on the scale $L=6$ is similar here and in the half-filled system with the same $K_2$ shown in the middle panel in Fig.~\ref{fig:GFMC_wavefunction_transition}; however, unlike the half-filled case, it does not renormalize upon increasing $L$ consistent with the picture where some relevant Umklapp becomes inoperative for $\rho < 1/2$.
(One needs to worry about higher-order Umklapps if the density happens to be commensurate, which we do not worry here, having more in mind incommensurate densities in a window $1/3 \lesssim \rho < 1/2$.  Density $\rho=1/3$ may be slightly outside the stability window if we take the VMC energy per site estimates in Fig.~\ref{fig:vmc_E_rho} seriously and is perhaps stabilized here against phase separation by the finite system size, but is a good example allowing us to see the absence of flow of the EBL stiffness with several our sizes.)
Similar result is obtained for $K_2=0.5$ (not shown), and both indicate that strong EBL signature is present.    In the bottom panel of Fig.~\ref{fig:GFMC_rho_cross}, the sharper ``cross'' shows that the EBL is more stable for density $\rho<1/2$ compared to the half-filled system.

We also note the presence of small $2k_F$ ridges (discussed in Appendix~\ref{app:EBLprecis}) which we typically observe in the GFMC density structure factor of uniform liquid ground states studied at densities away from half.    This feature is predicted in the EBL theory and may be taken as additional evidence for identifying the uniform phase with EBL.   Thus, our GFMC study for densities close to 1/2 shows that the intermediate $K_2$ regime is a stable EBL phase.   Together with the VMC results, we think the evidences are sufficiently strong to conclude that the EBL phase is already realized in the $K_1$-$K_2$ model for densities $1/3 \lesssim \rho < 1/2$ and intermediate $K_2$ values, while the phase separation dominates at lower density.

\begin{figure}
  \centering
  \includegraphics[width=\columnwidth]{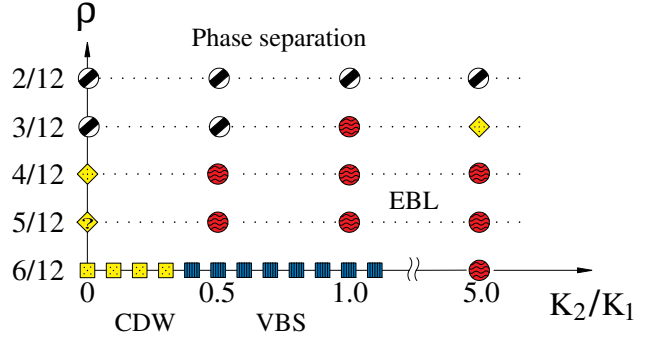}
  \caption{GFMC phase diagram for the $K_1$-$K_2$ model on the $12 \times 12$ lattice with varying boson density from 2 up to 6 bosons per row and per column.}
  \label{fig:completePD}
\end{figure}

Figure~\ref{fig:completePD} shows the extended GFMC phase diagram which includes densities below 1/2 that we studied.  We have also added points at $K_2 = 0$, where our similar study using the ``cross'' technique indicates that the EBL is unstable also close to half-filling (but we have not established the resulting phases).  Thus, it has been crucial to add moderate $K_2$ exchanges to realize the EBL away from half-filling.

\section{Conclusions}
\label{sec:conclusion}

In summary, we studied the $K_1$-$K_2$ hard-core boson model with ring-only exchanges and found a transition at half-filling from a staggered CDW order at small $K_2$ to a columnar VBS order at intermediate $K_2$, and tantalizingly realizing the EBL phase at large $K_2$.  For densities away from half-filling (but not too far), the EBL phase is more robust and our evidence strongly suggests that the EBL is already realized in this model for intermediate $K_2$.  (Having some $K_2$ is helpful, since the pure $K_1$ model does not appear to have the EBL also away from half-filling.)  For still lower densities, instability to phase separation is observed.

Although our sizes are significantly smaller than typically studied for more conventional phases with advanced techniques such as Stochastic Series Expansion, we already reach $12 \times 12$ systems which are much larger than sizes used in ED studies.  Like the ED, the GFMC method that we use provides exact information about the ground state and allows us to reasonably establish the phase diagram of the proposed model already with our sizes.  Our GFMC evidence of the EBL phase is quite suggestive at half filling, although we critically point out possible pitfalls.  Specifically, while direct measurements of the charge- or bond-solid order parameters do not reveal any orders, the detailed comparison of the long-wavelength signature in the density structure factor with the available EBL theory suggests that our tentative EBL points at half-filling are very close to being unstable.  (It may be helpful to modify the model still a bit like adding $2\times 2$ ring exchanges to more reliably stabilize the EBL in the half-filled system.)  On the other hand, similar comparison for the claimed EBL points away from half-filling indicates that they are safely away from instabilities.

We hope that it will be possible to study significantly larger sizes and confront our tentative EBL realizations as well as systematically confront the EBL theory.  There are a number of further properties and questions that one would like to explore.  Thus, we have measured only density and energy correlations that are diagonal in the number basis as they are simplest to implement in the GFMC.  The gaplessness of the EBL and its ``Excitonic'' character can be probed directly by measuring boson ``box'' correlations like $G_\phi^{(4)}(x,y) \equiv \la b_{(0,0)}^\dagger b_{(0,y)} b_{(x,y)}^\dagger b_{(x,0)} \ra$.  At fixed $y$, this can be viewed as an exciton propagator for excitons of size $y$, and is predicted to show power-law behavior $\sim |x|^{-\eta(y)}$ with calculable $y$-dependent exponent.\cite{Paramekanti2002}  A crucial characterization of the EBL (which, in particular, determines all power law exponents) is the ``EBL stiffness'' function.  While we have had some access to it via the cross analysis of the density structure factor, it would also be interesting to measure the EBL stiffness directly.  More broadly, the EBL is an example of a very gapless quantum liquid, and it can be challenging but fruitful learning grounds for how to handle such phases in Quantum Monte Carlo simulations.

One immediate question for the EBL theory is that in its present form it does not seem to anticipate the columnar VBS phase that we found numerically in the $K_1$-$K_2$ model.  Another question for both numerical and theoretical studies is to understand phases away from half-filling near $K_2 = 0$, in the regime where we did not find the phase separation but also concluded that the EBL is not stable.  

We also note that our present realizations of the EBL are very likely immediately unstable towards a superfluid if we allow unfrustrated boson hopping.  This is based on our estimates of the EBL phase stiffnesses and understanding of the EBL stability conditions against boson hopping.\cite{Paramekanti2002}  Thus, an EBL phase envisioned in Ref.~\onlinecite{Paramekanti2002} in such a broader sense [with no special conservation laws other than global U(1)] will not be realized with our $K_1$-$K_2$ model.  However, we hope that our work will stimulate further studies to achieve such a phase.

Our tentative numerical realization of the EBL (even in the restricted sense of ring-only systems) is of broader interest in the search for so-called Bose-metal phases or more generally non-Fermi liquid metals.\cite{Nagaosa2006, Motrunich2007, Kaul2008, Senthil2008, SSLee2009, Sheng2008, Fisher2008, Feiguin2010, Mross2010, Faulkner2010, Sachdev2010}  Despite many studies (including more recently of holographic metals in the High Energy Theory community\cite{Faulkner2010, Sachdev2010}), to date there is no example where such phases can be demonstrated controllably.  The EBL can be viewed as a special kind of a Bose-metal and its theory is on more firm ground (an interesting perspective on the EBL is that it can be viewed as a solvable example of a parton-gauge theory where partons have flat Fermi surfaces in meanfield, cf.\ Appendix~\ref{app:parton2EBL}).   It is hoped that the present work may trigger more refined studies of the EBL and confrontation with the theory, and more efforts to access the challenging but very interesting and topical Bose-metal phases.

\acknowledgments

We are thankful to M.~P.~A.~Fisher for inspiring discussions and encouragement throughout this work and R.~Kaul and A.~Paramekanti for discussions.  The research is supported by the NSF through grant DMR-0907145 and the A.~P.~Sloan Foundation.  We also acknowledge using the IT2 cluster at the Caltech CACR as well as the hospitality of the Max Planck Institute for the Physics of Complex Systems.

\appendix
\section{Precis of the EBL theory}
\label{app:EBLprecis}

For the benefit of readers, this Appendix summarizes results from Refs.~\onlinecite{Paramekanti2002, Xu2005, Xu2007, Balents2005} relevant for our numerical work, complementing them where needed.

\subsection{Gaussian fixed point theory}

The EBL phase is described by a fixed point Lagrangian\cite{Paramekanti2002}
\begin{eqnarray}
{\cal L}_0 &=& {\cal H}_0[\varphi, \vartheta] + \sum_\br \frac{i}{\pi} \partial_\tau \varphi(\br) \Delta_{xy} \vartheta(\bR) ~, \\
{\cal H}_0[\varphi, \vartheta] &=& \int_\bk \left[ \frac{\cK(\bk)}{2} |(\Delta_{xy} \varphi)_\bk|^2 + \frac{\cU(\bk)}{2\pi^2} |(\Delta_{xy} \vartheta)_\bk|^2 \right] ~. \nonumber
\end{eqnarray}
We use the same notation for the fields as in Ref.~\onlinecite{Paramekanti2002}: $\varphi$ corresponds to the spin wave part of the original phase variables, while $\vartheta$ describes coarse-grained boson density fluctuations $\delta n(\br) = n(\br) - \bar{n}$.  In the microscopic lattice derivation,\cite{Paramekanti2002}
\begin{eqnarray}
{\delta n}_0(\br) &=& \frac{1}{\pi} \Big[\vartheta(\bR) - \vartheta(\bR - \hbx) - \vartheta(\bR - \hby) \\
&&~~~ + \vartheta(\bR - \hbx - \hby) \Big] \equiv \frac{1}{\pi} \Delta_{xy} \vartheta ~. ~~~
\end{eqnarray}
Lower-case $\br = (x,y)$ denote the original lattice sites, while upper-case $\bR = (X,Y)$ denote the dual lattice sites, cf.\ Fig.~\ref{fig:dualvars}.  The above ${\cal H}_0$ contains general ``EBL parameters''\cite{Paramekanti2002} $\cK(\bk)$ and $\cU(\bk)$, which are functions of $\bk = (k_x, k_y)$ respecting square lattice symmetries.  Long-wavelength properties such as various power law exponents (discussed below) depend on the function $\cK(0, k_y)/\cU(0, k_y)$.  We will consider stability of this fixed point to allowed perturbations later.  A parton-gauge derivation utilizing familiar 1D Bosonization rules is sketched in Appendix~\ref{app:parton2EBL}.

Integrating out the field $\vartheta$ gives a spin wave theory Lagrangian in terms of the $\varphi$ field that corresponds to the spin wave Hamiltonian Eq.~(\ref{eq:hamiltonian_SW}).  However, the compactness of the original phase variables (or, equivalently, the discreteness of the boson number) is best captured in terms of the dual field $\vartheta$.  Therefore we consider a dual Lagrangian obtained upon integrating out the $\varphi$ field,
\begin{equation}
{\cal L}_{{\rm dual}, 0}[\vartheta] = \int_\bk \left[ \frac{\cU(\bk)}{2\pi^2} |(\Delta_{xy} \vartheta)_\bk|^2 + \frac{1}{2\pi^2 \cK(\bk)} |(\partial_\tau \vartheta)_\bk|^2 \right] ~.
\label{Ldual}
\end{equation}
Calculations mentioned below are performed in this Gaussian theory, and the main task is to express physical operators in terms of the continuum fields.

The Gaussian theory gives equal-time correlations of the coarse-grained boson density as\cite{Paramekanti2002}
\begin{eqnarray}
\la \big|\delta n_0(\bk) \big|^2 \ra &=& 2\; \sqrt{\frac{\cK(\bk)}{\cU(\bk)}} \big|\sin(k_x/2)\; \sin(k_y/2) \big| ~~~~~~ \\ 
&& \times |C_\rho(\bk)|^2 ~.
\label{cross}
\end{eqnarray}
The ``cross'' formed by singular lines $(0, k_y)$ and $(k_x, 0)$ is the most prominent feature in the density structure factor in the EBL phase.  Note that the result contains a non-universal factor $|C_\rho(\bk)|^2$, which is expected to arise after coarse-graining
${\delta n}_0(\br) \to \sum_\bfm C_\rho(\bfm) {\delta n}_0(\br + \bfm)$.
The coefficients $C_\rho(\bfm)$ are non-universal but satisfy $\sum_\bfm C_\rho(\bfm) = 1$, implying $C_\rho(\bk \to {\bf 0}) = 1$.  We conjecture that specification of the EBL fixed point theory (including identification of observables and residual interactions below) requires an even stronger condition: $C_\rho(k_x \to 0, k_y) = 1$ for any $k_y$.

A more accurate formula for the boson number operator contains also ``$2k_F$'' contributions (where $2k_F \equiv 2\pi\bar{n}$),
\begin{eqnarray}
n(\br) &=& \bar{n} + {\delta n}_0 + A \cos[2\nabla_y \vartheta + 2 \pi \bar{n} (x - 1/2)] ~~~~~~ \label{n2kFx} \\
&+& A \cos[2\nabla_x \vartheta + 2 \pi \bar{n} (y - 1/2)] ~,
\end{eqnarray}
where $\bar{n}$ is the boson density per site and
\begin{eqnarray}
\nabla_\mu \vartheta \equiv \vartheta(\bR) - \vartheta(\bR - \hat{\bm \mu}) ~.
\end{eqnarray}
[More precisely, the oscillatory contribution say in the first line in Eq.~(\ref{n2kFx}) should be written as $\sin[2\nabla_y \vartheta(X, y) + 2 \pi \bar{n} X] - \sin[2\nabla_y \vartheta(X-1, y) + 2 \pi \bar{n} (X-1)]$, where the original lattice coordinate $x$ is between dual lattice coordinates $X-1$ and $X$ as in Fig.~\ref{fig:dualvars}.  However, any fuzziness in the location of $\nabla_y \vartheta$ along the $\hbx$ direction tuns out to be unimportant and the more precise expression simplifies to that in Eq.~(\ref{n2kFx}).]   The particular oscillatory contributions are not explicitly listed in Ref.~\onlinecite{Paramekanti2002} but require a more sophisticated treatment there; they are discussed in Refs.~\onlinecite{Xu2005, Xu2007} and are also recapped in Appendix~\ref{app:parton2EBL} from the parton perspective.

To characterize the $2k_F$ oscillatory part of the boson density, we need correlations
\begin{equation}
\la e^{i 2\nabla_y \vartheta(\bR)} e^{-i 2\nabla_y \vartheta(\bR')} \ra \sim \frac{\delta_{Y,Y'}}{|X - X'|^{2\Delta_{2k_F}}} ~.
\label{2kFpwr}
\end{equation}
These are ultra-short-range in the $\hby$ direction.  The power law in the $\hbx$ direction (as well as in the imaginary time direction not written explicitly) is determined by the scaling dimension
\begin{equation}
\Delta_{2k_F} = 2 \int_0^\pi dk_y \sqrt{\frac{\cK(0, k_y)}{\cU(0, k_y)}} \; \sin(k_y/2) \doteq 4 \sqrt{\frac{\cK}{\cU}} ~.
\end{equation}
For an illustration, the very last equality specializes to the case of $\bk$-independent $\cK$ and $\cU$; here and below this case is marked with ``$\doteq$''.  Note that in the actual density correlations, $\delta_{Y,Y'}$ is replaced by an exponentially decaying function of $|Y-Y'|$, since by general symmetry reasoning we also expect contributions in Eq.~(\ref{n2kFx}) where $\nabla_y \vartheta$ is evaluated at $\br \pm j\hby$, albeit with exponentially decreasing amplitudes.  In $\bk$-space, the density structure factor has line singularities,
\begin{equation}
\sim |k_x \pm 2\pi\bar{n}|^{2\Delta_{2k_F} - 1} ~.
\label{2kFsing}
\end{equation}
In the parton perspective on the EBL, Appendix~\ref{app:parton2EBL}, these can be viewed as the $2k_F$ surfaces associated with partons moving in the $\hbx$ direction.  There are also similar $2k_F$ lines at $k_y = \pm 2\pi\bar{n}$.

Finally, the same singular lines are present in bond and plaquette correlators.  For example, for an energy-type operator associated with a bond $[\br, \br+\hbx]$, we have \cite{Paramekanti2002, Xu2005, Xu2007}
\begin{equation}
\delta {\cal B}_{\br, \br+\hbx} = A' \cos[2\nabla_y \vartheta + 2 \pi \bar{n} x] ~,
\label{B2kFx}
\end{equation}
which is similar to the oscillatory part in the density operator except for a phase shift.

\subsection{Nonlinear interactions}

The discreteness of the original boson charges can be faithfully represented by allowing non-linear terms in the $\vartheta$ variables.  The first such interactions that one would write down are\cite{Paramekanti2002}
\begin{equation}
{\cal L}_{\EuFrak{v}} = -\sum_{q=1}^\infty \EuFrak{v}_q \cos[q(2\vartheta + 2\pi\bar{n} X Y)] ~.
\label{vq}
\end{equation}
Such insertions have ultra-short-range correlations when calculated in the Gaussian action Eq.~(\ref{Ldual}), and small such residual interactions are therefore irrelevant in the EBL theory.  However, as discussed below, there are allowed additional interactions (which can be viewed as derived from the ${\cal L}_{\EuFrak{v}}$) that do have power law correlations, and then stability requires that the power laws are sufficiently fast.  Note that non-integer boson density causes oscillatory phases, which can effectively ``disallow'' interactions from the low-energy theory.  Note also that even though the terms in Eq.~(\ref{vq}) do not enter the stability condition for the EBL fixed point, once the EBL becomes unstable they are important to correctly describe resulting phases (see analysis in Ref.~\onlinecite{Paramekanti2002}, Sec.~V~B~2, and here in Sec.~\ref{subsubapp:instabilities}).

\subsection{Stability of the EBL at generic densities}
\label{subapp:gendens}

We first discuss stability of the EBL phase at incommensurate density $\bar{n}$.  For any $\bar{n}$, there is an allowed (non-oscillatory) interaction associated with chains parallel to $\hbx$ and separated by $j \hby$:
\begin{equation}
\mathcal{L}_{\EuFrak{h}} = -\EuFrak{h}^{(j \neq 0)} \cos\left[2 \nabla_y \vartheta(\bR) - 2 \nabla_y \vartheta(\bR + j\hby) \right] ~.
\label{hj}
\end{equation}
There are similar interactions associated with chains parallel to $\hby$.  Such interactions were not discussed in Ref.~\onlinecite{Paramekanti2002}, while $j=1$ case was considered in Ref.~\onlinecite{Xu2007} (see also Appendix~\ref{app:parton2EBL}).  The above operator has short-range correlations in the $\hby$ direction, but power law correlations in the $\hbx$ direction as well as in the imaginary time; the corresponding scaling dimension is
\begin{eqnarray}
\Delta[\EuFrak{h}^{(j)}] &=& 8 \int_0^\pi dk_y \sqrt{\frac{\cK(0, k_y)}{\cU(0, k_y)}} \; \sin^2(j k_y/2) \; \sin(k_y/2) \nonumber \\
&\doteq& 8 \left(1 + \frac{1}{4j^2 - 1}\right) \sqrt{\frac{\cK}{\cU}} ~.
\end{eqnarray}
The last line specializes to the case of $\bk$-independent $\cK$ and $\cU$; some representative values are
\begin{eqnarray}
\Delta[\EuFrak{h}^{(j=\pm 1)}] &\doteq& \frac{32}{3} \sqrt{\frac{\cK}{\cU}} \;\approx\; 10.6666 \sqrt{\frac{\cK}{\cU}} ~, \\
\Delta[\EuFrak{h}^{(j=\pm 2)}] &\doteq& \frac{128}{15} \sqrt{\frac{\cK}{\cU}} \;\approx\; 8.5333 \sqrt{\frac{\cK}{\cU}} ~, \\
\Delta[\EuFrak{h}^{(j=\infty)}] &\doteq& 8 \sqrt{\frac{\cK}{\cU}} ~.
\end{eqnarray}
Interestingly, in this case the smallest scaling dimension is achieved for $j \to \infty$, probably because of particular inter-chain correlations present in the system.  

Without any assumptions on the parameters $\cK(\bk)$ and $\cU(\bk)$, the scaling dimension in the $j \to \infty$ limit can be related to the $2k_F$ correlation dimension:
\begin{equation}
\Delta[\EuFrak{h}^{(j=\infty)}] = 2 \Delta_{2k_F} ~.
\end{equation}
References~\onlinecite{Paramekanti2002, Xu2005, Xu2007} argued that stability to such residual interactions is determined by comparing the scaling dimensions to a reduced space-time dimensionality equal to $1 + 1 = 2$ here.  Adopting this criterion, we conclude that stability requires $\Delta_{2k_F} > 1$.  Therefore, the $2k_F$ correlations must decay faster than $1/x^2$ along chains, or, equivalently, the corresponding singularity Eq.~(\ref{2kFsing}) across the $2k_F$ lines in the momentum space must be weaker than slope discontinuity.

It is not clear at present what happens if the system is unstable to the above interactions in the incommensurate case, i.e., in the absence of Umklapp terms.
Another remark is that if one also allows boson hopping between chains and considers $\bk$-independent $\cK$ and $\cU$, then the scaling dimension for the boson hopping works out\cite{Paramekanti2002} to be such that there is no stable EBL phase with unfrustrated hopping even at incommensurate density.  It would be interesting to explore different EBL stiffness functions $\sqrt{\cK(0, k_y) / \cU(0, k_y)}$ that may give stability against both $\mathcal{L}_{\EuFrak{h}}$, Eq.~(\ref{hj}), and boson hopping.  In the present paper, we focus solely on the restricted class of models with no boson hopping.

\subsection{Stability of the EBL at half-filling}

Let us consider stability at density $\bar{n} = 1/2$ where large part of our numerical work is performed.  In this case, the following interactions are also allowed,
\begin{equation}
\mathcal{L}_{\EuFrak{u}} = -\EuFrak{u}^{(j)} \cos\left[2 \nabla_y \vartheta(\bR) + 2 \nabla_y \vartheta(\bR + j\hby) \right] ~.
\label{uj}
\end{equation}
Here $j=0$ corresponds to an Umklapp in each chain, while general $j$ corresponds to an Umklapp involving chains separated by $j \hby$ (cf.\ Appendix~\ref{app:parton2EBL}).  There are similar interactions associated with chains parallel to $\hby$.  The corresponding scaling dimensions, in the same sense as before, are
\begin{eqnarray}
\Delta[\EuFrak{u}^{(j)}] &=& 8 \int_0^\pi dk_y \sqrt{\frac{\cK(0, k_y)}{\cU(0, k_y)}} \; \cos^2(j k_y/2) \; \sin(k_y/2) \nonumber \\
&\doteq& 8 \left(1 - \frac{1}{4 j^2 - 1}\right) \sqrt{\frac{\cK}{\cU}} ~.
\end{eqnarray}
The last line again specializes to the case of $\bk$-independent $\cK$ and $\cU$.  Here the smallest scaling dimension is obtained for $j = \pm 1$ corresponding to interactions between nearest neighbor chains,
\begin{eqnarray}
\Delta[\EuFrak{u}^{(j = \pm 1)}] &\doteq& \frac{16}{3} \; \sqrt{\frac{\cK}{\cU}} \;\approx\; 5.3333 \sqrt{\frac{\cK}{\cU}} ~, \label{u1pwr} \\
\Delta[\EuFrak{u}^{(j = \pm 2)}] &\doteq& \frac{112}{15} \; \sqrt{\frac{\cK}{\cU}} \;\approx\; 7.4666 \sqrt{\frac{\cK}{\cU}} ~, \\
\Delta[\EuFrak{u}^{(j = 0)}] &\doteq& 16 \; \sqrt{\frac{\cK}{\cU}} ~.
\end{eqnarray}
The last two lines give the scaling dimensions of the second-neighbor-chain Umklapp and the intra-chain Umklapp.

Reference~\onlinecite{Paramekanti2002} considered the following Umklapps at half-filling,
\begin{equation}
\mathcal{L}_{\EuFrak{w}} = -\EuFrak{w}_{2q, m} \cos\left[2 q \vartheta(\bR) - 2 q \vartheta(\bR + m\hby) \right] ~,
\label{wm}
\end{equation}
with the condition that $q m$ is an even integer.  We have $\EuFrak{w}_{2, 2} = \EuFrak{u}^{(1)}$ and $\EuFrak{w}_{4, 1} = \EuFrak{u}^{(0)}$, but in general the sets $\EuFrak{u}^{(j)}$ and $\EuFrak{w}_{2q, m}$ are different.  The latter Umklapps can be viewed as involving $m$ neighboring chains (cf.\ Appendix~\ref{app:parton2EBL}).  The scaling dimensions are\cite{Paramekanti2002}
\begin{eqnarray}
\Delta[\EuFrak{w}_{2q, m}] &=& 2 q^2 \int_0^\pi dk_y \sqrt{\frac{\cK(0, k_y)}{\cU(0, k_y)}} \; \frac{\sin^2(m k_y/2)}{\sin(k_y/2)} \nonumber \\
&\doteq& 4 q^2 \left(1 + \frac{1}{3} + \frac{1}{5} + \dots + \frac{1}{2m - 1} \right) \sqrt{\frac{\cK}{\cU}} ~, \nonumber 
\end{eqnarray}
where the last line specializes to the case of $\bk$-independent $\cK$ and $\cU$.
We also caution that more complex interactions not considered here may get reduced scaling dimensions for some functional forms of $\cK(0, k_y)/\cU(0, k_y)$.

As far as boson ring models are concerned, the EBL fixed point is stable for sufficient dominance of the effective ``ringing'' $\cK$ over the ``repulsion'' $\cU$.

\subsubsection{Instabilities at half-filling}
\label{subsubapp:instabilities}

Let us consider the case when the most relevant term is the nearest-chain Umklapp $\EuFrak{w}_{2, 2} = \EuFrak{u}^{(1)}$ [this is true, e.g., for $\bk$-independent $\cK/\cU$, see Eq.~(\ref{u1pwr})].  We also assume $\EuFrak{u}^{(1)} > 0$, which is expected in the presence of effective nearest-neighbor repulsion, see discussion after Eq.~(\ref{Vj}).

It is convenient to define
\begin{eqnarray}
\theta_1(X, y) &\equiv& \nabla_y \vartheta ~, \\
\theta_2(x, Y) &\equiv& \nabla_x \vartheta ~.
\end{eqnarray}
In principle, all discussion below can be carried out using $\vartheta$ variables only and is essentially equivalent to the analysis in Ref.~\onlinecite{Paramekanti2002}, Sec.~V~B~2; one addition is discussion of interactions and regimes that can produce a pure CDW state.  Nevertheless, the $\theta_1$ and $\theta_2$ variables introduced here make a connection with the parton-gauge perspective on the EBL, Appendix~\ref{app:parton2EBL}, and, up to a point, allow to use quasi-one-dimensional language and intuition.  For example, the interactions Eqs.~(\ref{hj}) and (\ref{uj}) are simply expressed in terms of the $\theta_1$ or $\theta_2$ and are naturally associated with parallel chains in one or the other direction.  However, note that these variables are constrained by $\nabla_x \theta_1 = \nabla_y \theta_2$.

When the $\EuFrak{u}^{(1)}$ Umklapp is relevant and flows to strong coupling, it provides pinning
\begin{eqnarray}
2\theta_1(X, y) &=& - 2\theta_1(X, y-1) = (-1)^y \, 2\theta_1(X, 0) ~~~~~ \\
&=& (-1)^y \, 2\theta_1(0,0) ~, \\
2\theta_2(x, Y) &=& (-1)^x \, 2\theta_2(0,0) ~,
\end{eqnarray}
where all equalities are modulo $2\pi$, which is natural periodicity for the $2\vartheta$, $2\theta_1$, and $2\theta_2$ variables.  In the above, we have also minimized the quadratic energy Eq.~(\ref{Ldual}) by looking for time-independent fields satisfying $\nabla_x \theta_1 = \nabla_y \theta_2 = 0$.  For the $\vartheta$ variable we obtain
\begin{eqnarray}
2\vartheta(X, Y) &=& 2\vartheta(0,0) + \frac{(-1)^y - 1}{2} \, 2\theta_1(0,0) ~~ \\
&+& \frac{(-1)^x - 1}{2} \, 2\theta_2(0,0) ~.
\label{threeparams}
\end{eqnarray}
We are thus left with three undetermined parameters $2\vartheta(0,0)$, $2\theta_1(0,0)$, and $2\theta_2(0,0)$.

We can loosely view the nearest-chain Umklapp $\EuFrak{u}^{(1)}$ as locking relative density fluctuations in neighboring chains but not locking the density to the lattice.  The latter can be provided by several terms which we list roughly in the order of their relevance in the EBL theory and which potentially have large bare values: second-chain Umklapp $-\EuFrak{u}^{(2)} \cos[2\theta_1(X, y) + 2\theta_1(X, y+2)]$; near-chain non-Umklapp $-\EuFrak{h}^{(1)} \cos[2\theta_1(X, y) - 2\theta_1(X, y+1)]$; and intra-chain Umklapp $-\EuFrak{u}^{(0)} \cos[4\theta_1(X, y)]$.  We have only shown terms associated with chains running in the $\hbx$ direction; there are also similar terms associated with chains in the $\hby$ direction.  Putting everything together, we obtain an effective pinning potential on the $\theta_1(0,0)$ and $\theta_2(0,0)$,
\begin{equation}
\label{lambdaeff}
-\lambda_{\rm eff} \Big\{\cos[4 \theta_1(0,0)] + \cos[4 \theta_2(0,0)] \Big\} ~.
\end{equation}
Schematically, ignoring any intermediate scale RG flows,
\begin{equation}
\lambda_{\rm eff} = \EuFrak{u}^{(2)} + \EuFrak{h}^{(1)} + \EuFrak{u}^{(0)} ~.
\end{equation}

Let us first consider the case $\lambda_{\rm eff} > 0$.  This gives pinning
\begin{eqnarray}
2 \theta_1(0,0) &=& \pi m_1 ~,\\
2 \theta_2(0,0) &=& \pi m_2 ~,
\label{thtpin4plaquette}
\end{eqnarray}
where integers $m_1, m_2$ can take independent values $0$ and $1$.  By examining observables Eq.~(\ref{B2kFx}), we can already conclude that the system has columnar dimer patterns on the $x$- and $y$-bonds:
\begin{eqnarray}
\delta {\cal B}_{\br, \br+\hbx} &=& A' (-1)^x \cos[2\theta_1(X, y)] \\
&=& A' (-1)^x (-1)^{m_1} ~, \\
\delta {\cal B}_{\br, \br+\hby} &=&  A' (-1)^y (-1)^{m_2} ~.
\end{eqnarray}
The simultaneous presence of these bond energy patterns corresponds to a plaquette solid.  To complete the analysis, we also need to consider pinning of the $2\vartheta(0,0)$.  This is provided by terms in Eq.~(\ref{vq}).  For simplicity, consider just the $\EuFrak{v}_1$ term.  Using Eqs.~(\ref{threeparams}) and (\ref{thtpin4plaquette}), this energy is proportional to $-\EuFrak{v}_1 \cos[2\vartheta(0,0)] [1 + (-1)^{m_1} + (-1)^{m_2} - (-1)^{m_1 + m_2}]$.  We readily establish that for each pair $\{m_1, m_2\}$, the $2\vartheta(0,0)$ is uniquely determined and gives a state where three out of four plaquettes have more negative energies, i.e., a plaquette solid; four distinct pairs $\{m_1, m_2\}$ correspond to four ways to put this solid on the lattice.
(Original Ref.~\onlinecite{Paramekanti2002} also considered adding $\EuFrak{v}_2$ to the energy mix and argued that for sufficiently negative $\EuFrak{v}_2$ the system will have coexisting plaquette and CDW orders.)

Let us now consider the case $\lambda_{\rm eff} < 0$.  This gives pinning
\begin{eqnarray}
2 \theta_1(0,0) &=& \frac{\pi}{2} + \pi m_1 ~,\\
2 \theta_2(0,0) &=& \frac{\pi}{2} + \pi m_2 ~.
\label{thtpin4CDW}
\end{eqnarray}
By examining observables Eq.~(\ref{n2kFx}), we can already conclude that the system has $(\pi,\pi)$ CDW order:
\begin{eqnarray}
\delta n_1(\br) &=& A (-1)^x \sin[2\theta_1(X, y)] \\
&=& A (-1)^{x+y} (-1)^{m_1} ~, \\
\delta n_2(\br) &=& A (-1)^{x+y} (-1)^{m_2} ~. 
\end{eqnarray}
To be more precise, we also need to consider the pinning of the $2\vartheta(0,0)$ by terms in Eq.~(\ref{vq}).  Using Eqs.~(\ref{threeparams}) and (\ref{thtpin4CDW}), the $\EuFrak{v}_1$ energy is proportional to $-\EuFrak{v}_1 \{ \cos[2\vartheta(0,0)] [1 + (-1)^{m_1 + m_2}] + \sin[2\vartheta(0,0)] [(-1)^{m_1} + (-1)^{m_2}] \}$.  This is minimized when $m_1 = m_2$ and unique $2\vartheta(0,0)$ in each case, and produces uniform plaquette energies.  The two independent minima $m_1 = m_2 = 0$ or $1$ correspond to two ways of putting the CDW on the lattice.  [A note on approaches:  Our Eq.~(\ref{n2kFx}) assumes some coarse-graining, i.e., some high-energy fields are already integrated out.  We could instead proceed using bare lattice variables $\vartheta$ where $\delta n(\br) = \Delta_{xy} \vartheta/\pi$ and minimize the dual action including the nonlinear terms $\EuFrak{u}_1, \EuFrak{u}_2, \EuFrak{h}_1, \EuFrak{u}_0, \EuFrak{v}_1$, etc.  We would then find that a careful treatment of the saddle point approximated by Eq.~(\ref{thtpin4CDW}) indeed has a staggered charge density wave, $\delta n(\br) \sim (-1)^{x+y}$.]

At this stage, we can speculate that the ring model at $K_2 = 0$ has small bare value of $\EuFrak{u}_2$ and large bare values $\EuFrak{h}_1 < 0, \EuFrak{u}_0 < 0$ due to repulsive effective nearest-neighbor interactions, see discussion after Eq.~(\ref{Vj}).  In this case, $\lambda_{\rm eff} < 0$ and the CDW phase is realized.  As we increase $K_2$, it likely feeds directly into bare $\EuFrak{u}_2 > 0$ because of the effective second-neighbor avoidance desired by such ring terms.  This can eventually make $\lambda_{\rm eff} > 0$ and change the order to plaquette solid.  However, our numerics suggests that the $K_1$-$K_2$ model at intermediate $K_2$ realizes the columnar VBS instead.  We have not succeeded to understand this within the above EBL instability treatments.  It would be good to clarify this since the two bond-solids are usually related\cite{ReadSachdev1989, Lannert2001, Senthil2004science, Senthil2004} and perhaps we are missing some physics ingredients.  [For example, one can contemplate more interactions, say leading to terms like $-\lambda' \cos[4 \theta_1(0,0)] \cos[4 \theta_2(0,0)]$ in addition to Eq.~(\ref{lambdaeff}), which can indeed produce columnar VBS but only coexisting with the CDW.]

\section{Parton-gauge perspective on the EBL}
\label{app:parton2EBL}

Here we offer a parton-gauge perspective on the EBL phase.  An effective ``dimensional reduction'' noted by previous authors\cite{Paramekanti2002, Xu2005, Xu2007, Nussinov2005, Batista2005, Nussinov2006} can be also related to the one-dimensional (1D) character of partons.  The partons are still strongly interacting, but here we can treat all gauge fluctuation effects accurately and in fact arrive at the EBL description summarized in Appendix~\ref{app:EBLprecis}.  We will also see how the familiar 1D Bosonization techniques\cite{Haldane1981, Fisher1989} allow to quickly obtain physical observables and important residual interactions in the EBL theory.  [We emphasize, however, that the EBL phase is qualitatively different from sliding or crossed-sliding Luttinger liquids\cite{Emery2000, Vishwanath2001, Ranjan2001short, Ranjan2001long}---for example, it has specific heat $C \sim T \ln(1/T)$, cf.\ Ref.~\onlinecite{Paramekanti2002}.]

We write
\begin{equation}
b^\dagger(\br) = b_1^\dagger(\br) \, b_2^\dagger(\br)
\end{equation}
and recover the physical Hilbert space by the constraint
\begin{equation}
n(\br) = n_1(\br) = n_2(\br) ~.
\label{n1n2n}
\end{equation}
We will arrive at the EBL theory by starting from a ``mean field'' where $b_1$ partons hop only in the $\hbx$ direction while $b_2$ hop only in the $\hby$ direction.\cite{Motrunich2007} One can justify such a starting point, e.g., by noting that this mean field gives large negative boson ring energies.  As is familiar in slave particle treatments, a theory of fluctuations beyond the mean field contains a gauge field $a_x(\br), a_y(\br)$ residing on the links of the lattice; an alternative route connecting the parton-gauge system and bosonic ring model can be found in Sec.~IIIA of Ref.~\onlinecite{Motrunich2007}.  Here the partons $b_1$ and $b_2$ carry opposite gauge charges with respect to ${\bm a} = (a_x, a_y)$.  A complete theory also needs to treat the constraints Eq.~(\ref{n1n2n}).  Typically this would be done by introducing an auxiliary field which would then be interpreted as a temporal component of the gauge field.  However, here we are able to work with the constraints without the need for such a new field.

\begin{figure}[t]
\includegraphics[width=2.5in]{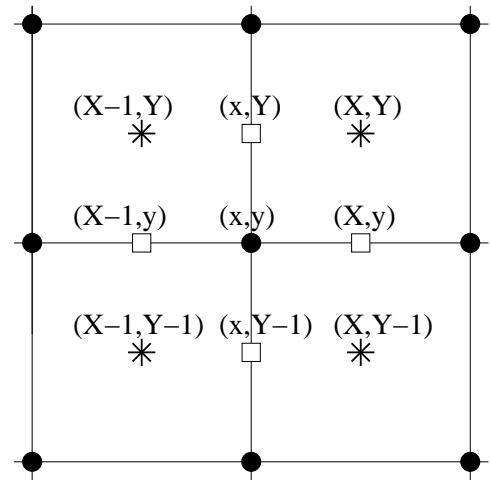}
\caption{
The parton phase fields $\varphi_1(x, y), \varphi_2(x, y)$ reside on the sites of the original square lattice (black circles); the 1D dual field $\theta_1(X, y)$ resides on the horizontal links and $\theta_2(x, Y)$ on the vertical links (white boxes).  The dual EBL theory field $\vartheta(X, Y)$ resides on the plaquettes of the original lattice, or equivalently on the sites of the dual lattice (stars). 
}
\label{fig:dualvars}
\end{figure}

We use phase and dual variables $\varphi_1(x, y)$ and $\theta_1(X, y)$ to describe a harmonic fluid\cite{Haldane1981, Fisher1989} of $b_1$ partons moving on a horizontally oriented chain located at a vertical coordinate $y$.  Similarly, we use variables $\varphi_2(x, y)$ and $\theta_2(x, Y)$ to describe a harmonic fluid of $b_2$ partons moving on a vertically oriented chain at a horizontal coordinate $x$.  Note that $\varphi_1(x, y)$ and $\varphi_2(x, y)$ reside on the sites of the original lattice, while $\theta_1(X, y)$ resides on the horizontal links and $\theta_2(x, Y)$ on the vertical links as illustrated in Fig.~\ref{fig:dualvars}.  Including coupling of the partons to the gauge field, the Gaussian part of the imaginary time Lagrangian reads
\begin{eqnarray}
\label{Lparton}
&&{\cal L}[\varphi_1, \theta_1, \varphi_2, \theta_2, a_x, a_y] = \\
&& = \sum_\br \Big[\frac{J}{2} (\nabla_x \varphi_1 - a_x)^2 + \frac{u}{2} \left(\frac{\nabla_x \theta_1}{\pi}\right)^2 + \frac{i}{\pi} \partial_\tau \varphi_1 \nabla_x \theta_1 \Big] \nonumber \\
&& + \sum_\br \Big[\frac{J}{2} (\nabla_y \varphi_2 + a_y)^2 + \frac{u}{2} \left(\frac{\nabla_y \theta_2}{\pi}\right)^2 + \frac{i}{\pi} \partial_\tau \varphi_2 \nabla_y \theta_2 \Big] \nonumber \\
&& + \sum_\br \frac{\kappa}{2} (\nabla_x a_y - \nabla_y a_x)^2 ~. \nonumber
\end{eqnarray}
Here for simplicity we showed the nearest-neighbor ``parton hopping'' coupling $J$ and on-site ``repulsion'' $u$, but all analysis below can be carried out more generally.  Crucially, we assume a stable phase where $(a_x, a_y)$ can be treated as a non-compact gauge field with a large ``stiffness'' parameter $\kappa$ (this assumes strong energetics selection of the particular mean field state by the ring exchanges of the microscopic boson model).  The parton densities are
\begin{eqnarray}
n_1(x, y) &=& \frac{\nabla_x \theta_1}{\pi} ~, \\
n_2(x, y) &=& \frac{\nabla_y \theta_2}{\pi} ~,
\end{eqnarray}
so the constraint Eq.~(\ref{n1n2n}) reads
\begin{equation}
\nabla_x \theta_1 = \nabla_y \theta_2 ~.
\label{tht-constraint}
\end{equation}

At this stage, we can integrate out the fields $\varphi_1$ and $\varphi_2$ and obtain Lagrangian density
\begin{eqnarray*}
&& \EuFrak{l}[\theta_1, \theta_2, a_x, a_y] = \frac{u}{2\pi^2}\left[ (\nabla_x \theta_1)^2 + (\nabla_y \theta_2)^2 \right] \\
&& ~~~~ + \frac{1}{2\pi^2 J} \left[ (\partial_\tau \theta_1)^2 + (\partial_\tau \theta_2)^2 \right] \\
&& ~~~~ + \frac{i}{\pi} (a_x\, \partial_\tau \theta_1 - a_y\, \partial_\tau \theta_2) + \frac{\kappa}{2} (\nabla_x a_y - \nabla_y a_x)^2 ~.
\end{eqnarray*}
Next, we solve the constraint Eq.~(\ref{tht-constraint}) via
\begin{eqnarray}
\theta_1(X, y) = \nabla_y \vartheta \equiv \vartheta(X, Y) - \vartheta(X, Y-1) ~, \label{tht1} \\
\theta_2(x, Y) = \nabla_x \vartheta \equiv \vartheta(X, Y) - \vartheta(X-1, Y) ~, \label{tht2}
\end{eqnarray}
cf.~Fig.~\ref{fig:dualvars}.  Integrating out the field ${\bm a}$, we finally obtain Lagrangian density
\begin{eqnarray}
\EuFrak{l}[\vartheta] = \frac{u}{\pi^2} (\nabla^2_{xy} \vartheta)^2 + \frac{1}{2\pi^2 J} (\partial_\tau {\bm \nabla} \vartheta)^2 + \frac{1}{2\pi^2 \kappa} (\partial_\tau \vartheta)^2 .
\end{eqnarray}
This is essentially the EBL theory written in the dual variables $\vartheta$, Eq.~(\ref{Ldual}), with $\cU(\bk) = 2u$ and $1/\cK(\bk) = 4[\sin^2(k_x/2) + \sin^2(k_y/2)]/J + 1/\kappa$.  [Note that one may be tempted to drop the $(\partial_\tau {\bm \nabla} \vartheta)^2 / J$ term as it contains more derivatives than the $(\partial_\tau \vartheta)^2 / \kappa$ term.  However, the long-distance EBL properties such as power-law exponents are determined by the full function $\cK(0, k_y)$ which does depend on $J$ if we want to be accurate in the simple model Eq.~(\ref{Lparton}) that we took.]  If we include from the start general interactions among the partons and general Maxwell terms for the gauge field, we obtain the general Gaussian EBL theory with $\bk$-dependent $\cK(\bk)$, $\cU(\bk)$ described in Appendix~\ref{app:EBLprecis}.

We can now establish connections between microscopic observables and the EBL theory in the $\vartheta$ variables.  Thus, in the 1D Bosonization treatment, the particle density and bond energy are given by
\begin{eqnarray}
\delta n_1(\br) &=& \frac{\nabla_x \theta_1}{\pi} + A \cos\left[2\theta_1 + 2\pi \bar{n}(x - 1/2)\right] ~,~~ \\
\delta B_{\br, \br+\hbx} &=& A' \cos\left[2\theta_1 + 2\pi \bar{n} x \right] ~.
\end{eqnarray}
Here $\theta_1$ already means the long-wavelength component and the precise location where it is evaluated along the chain is unimportant.  Writing $\theta_1$ via Eq.~(\ref{tht1}) we obtain Eqs.~(\ref{n2kFx}) and (\ref{B2kFx}) quoted in the EBL theory precis.

We can also express inter-chain density-density interactions
\begin{eqnarray}
V_j n_1(\br) n_1(\br + j\hby) \sim
V_j \cos\left[2\theta_1(\br) - 2\theta_1(\br + j\hby) \right] ~~ \\
+ V_j \cos\left[2\theta_1(\br) + 2\theta_1(\br + j\hby) + 4\pi\bar{n} x - 2\pi\bar{n} \right] ~, ~~
\label{Vj}
\end{eqnarray}
where we have only retained cosine terms.  Written in terms of the $\vartheta$ fields, the first line corresponds to the non-Umklapp interaction Eq.~(\ref{hj}) with $\EuFrak{h}^{(j)} = -V_j$.  The second line is non-oscillatory only at half-filling and corresponds to the Umklapp interaction Eq.~(\ref{uj}) with $\EuFrak{u}^{(j)} = V_j$.  Finally, terms of the type Eq.~(\ref{wm}) at half-filling with $q = 1$, $m = {\rm even}$, arise from Umklapps like $\cos\left[\sum_{j=1}^m 2 \theta_1(\br + j\hby) \right]$.

Let us remark about the effects of compactness of the microscopic gauge field.  As is known from the 1D folklore, allowing cosines of the dual fields in the action effectively allows vortices in the microscopic phase variables and provides a faithful treatment of the compactness of the phase variables.  In this respect one may wonder about the status of our theory once we allow the described cosine terms in the $\theta_1$ and $\theta_2$ variables.  It turns out that it is not complete yet, but becomes so after we allow terms like Eq.~(\ref{vq}) which in the parton-gauge setup correspond to allowing monopoles in the microscopic gauge field.  Since the insertions Eq.~(\ref{vq}) have ultra-short-range correlations at the EBL fixed point,\cite{Paramekanti2002} the issue of monopoles can be safely ignored in the stable EBL theory (but of course they cannot be ignored if the EBL becomes unstable and the parton fields acquire gaps).

We also note that while we used bosonic partons, a theory similar to that in Eq.~(\ref{Lparton}) would arise for a so-called extremal DLBL state of Ref.~\onlinecite{Motrunich2007}, where on the mean field level one starts with fermionic partons with flat Fermi surfaces (in the context of ring models with $K_1 < 0$).  Thus, the EBL can be viewed as a solution of such a special parton-gauge system, which nevertheless already has remarkable properties such as the non-Fermi-liquid specific heat\cite{Paramekanti2002} $C \sim T \ln(1/T)$.


\bibliography{biblio}

\end{document}